\documentclass{emulateapj}
\usepackage{apjfonts}

\shorttitle{AGN EVOLUTION IN CLUSTERS}
\shortauthors{MARTINI, SIVAKOFF, \& MULCHAEY}
\slugcomment{ApJ accepted [30 May 2009]}

\newcommand{\um}{\mu{\rm m}}

\newcommand{\eg}{{\rm e.g.}}

\newcommand{\ergs}{erg s$^{-1}$}
\newcommand{\kms}{km s$^{-1}$}
\newcommand{\chandra}{{\it Chandra}}
\newcommand{\spitzer}{{\it Spitzer}}
\newcommand{\oiii}{[\ion{O}{3}]}
\newcommand{\oii}{{\rm [\ion{O}{2}]}}

\newcommand{\msun}{M$_\odot$}

\begin{document}

\title{The Evolution of Active Galactic Nuclei in Clusters of Galaxies to 
Redshift 1.3} 

\author{Paul Martini, Gregory R. Sivakoff\altaffilmark{1}} 

\affil{Department of Astronomy and Center for Cosmology and Astroparticle 
Physics, The Ohio State University, 140 West 18th Avenue, Columbus, OH 43210, 
martini@astronomy.ohio-state.edu}

\altaffiltext{1}{Current Address: Department of Astronomy, University of Virginia, P.O. Box 400325, Charlottesville, VA 22904-4325}

\author{John S. Mulchaey} 

\affil{Carnegie Observatories, 813 Santa Barbara St., Pasadena, CA 91101-1292} 

\begin{abstract}

We have measured the luminous AGN population in a large sample of clusters of 
galaxies and find evidence for a substantial increase in the cluster AGN 
population from $z\sim0.05$ to $z\sim1.3$. The present sample now includes 
32 clusters of galaxies, including 15 clusters above $z=0.4$, which corresponds 
to a three-fold increase compared to our previous work at high redshift. 
At $z<0.4$ we have obtained new observations of AGN candidates in six 
additional clusters and found no new luminous AGN in cluster members. Our 
total sample of 17 low-redshift clusters contains only two luminous AGN, while 
at high redshifts there are 18 such AGN, or an average of more than one per 
cluster. We have characterized the evolution of luminous X-ray AGN as the 
fraction of galaxies with $M_R < M_R^*(z)+1$ that host AGN with rest-frame, 
hard X-ray [2--10 keV] luminosities $L_{X,H} \geq 10^{43}$ \ergs. The AGN 
fraction increases from $f_A = 0.134^{+0.18}_{-0.087}$\% at a median $z=0.19$ 
to $f_A = 1.00^{+0.29}_{-0.23}$\% at a median $z=0.72$. Our best estimate of 
the evolution is a factor of eight increase to $z=1$ and the statistical 
significance of the increase is $3.8\sigma$. This dramatic evolution is 
qualitatively similar to the evolution of the star-forming galaxy population 
in clusters known as the Butcher-Oemler effect. We discuss the implications of 
this result for the coevolution of black holes and galaxies in clusters, the 
evolution of AGN feedback, searches for clusters with the Sunyaev-Zel'dovich 
effect, and the possible detection of environment-dependent downsizing. 

\end{abstract}

\keywords{galaxies: active -- galaxies: clusters: general 
-- galaxies: evolution -- X-rays: galaxies -- X-rays: galaxies: clusters -- 
X-rays: general
}

\section{Introduction} \label{sec:intro} 

The demographics of AGN in clusters of galaxies have important implications 
for the growth of the supermassive black holes at the centers of 
cluster galaxies, the nature of AGN fueling, and the impact of AGN on the 
intracluster medium (ICM) over cosmic time. The luminous, massive elliptical 
galaxies that dominate the galaxy population in the richest clusters are 
also expected \citep[and in some cases are measured:][]{houghton06,gebhardt07} 
to have the most massive black holes in the local universe. As the 
stars in these galaxies appear to have an earlier mean formation epoch 
than those in field galaxies \citep[e.g.][]{vandokkum96,kelson97}, the apparent 
coevolution of black holes and galaxies \citep[e.g.][and references 
therein]{hopkins06a} implies that the bulk of their present black hole 
mass was also accreted at earlier times. 

This scenario is also motivated by observations of local clusters 
that clearly show their galaxy populations are more quiescent than 
local field galaxies. An early demonstration by \citet{osterbrock60} showed 
that cluster ellipticals were far less likely to have \oii\ $\lambda3727$ 
emission than field ellipticals, a result that has since been confirmed by many 
studies \citep[e.g.][]{gisler78,dressler85,dressler99}. One big question 
that has motivated this work is: Why are galaxy populations 
different in clusters? Numerous physical mechanisms have been invoked to 
explain the relative lack of star formation in cluster galaxies, as well as 
their higher fraction of elliptical and S0 galaxies \citep{dressler80} and 
relative lack of cold gas \citep[e.g.][]{giovanelli85}. These include 
ram-pressure stripping by the ICM \citep{gunn72}, 
evaporation of a galaxy's 
interstellar medium (ISM) by the hot ICM \citep{cowie77}, tidal effects 
with the cluster potential \citep{farouki81,merritt83a,byrd90}, the absence 
of newly-accreted cold gas \citep{larson80}, and galaxy harassment 
and mergers \citep{richstone76,moore96}. 

All of these physical effects may also be important for fueling accretion 
onto the central black holes in galaxies because they impact either the
available gas supply in a galaxy, angular momentum transport, or both. 
The best and perhaps only candidate process for fueling the most luminous AGN 
is the merger of two gas-rich galaxies \citep[e.g.][]{barnes92} and the 
relative lack of both cold gas and major mergers is a reasonable explanation 
for the nearly complete absence of QSOs hosted by cluster galaxies. For
less luminous AGN the case is less clear because an increasing number of 
physical processes such as minor mergers, galaxy harassment, various 
types of bars, stellar mass loss, etc.\ could also play a role 
\citep[see][for a review]{martini04c}. If mechanisms such as galaxy harassment 
and stellar mass loss are important for fueling low-luminosity AGN, then 
comparable numbers of low-luminosity AGN may be present in clusters and 
the field. 

Recent studies of the AGN fraction as a function of environment with 
emission-line galaxies from the Sloan Digital Sky Survey (SDSS) 
find that the most luminous AGN are rarer in denser environments
\citep[SDSS;][]{kauffmann04,popesso06}, although these studies do 
not sample the densest regions of clusters well. This decrease is in contrast 
to both lower-luminosity AGN in SDSS \citep{miller03a} and radio observations 
\citep{best04,best05b}, which show that the radio AGN fraction does not 
decrease significantly in denser environments. 
X-ray observations with \chandra\ show that the X-ray AGN fraction is 
larger than expected from AGN selection via visible-wavelength emission-lines. 
In previous work we showed that X-ray observations identified approximately 
five times as many AGN as selection at visible-wavelengths \citep{martini02,
martini06}, 
although the precise value of the X-ray excess depends significantly on the 
relative sensitivity and luminosity threshold of the observations. 
This spectroscopic study of X-ray counterparts confirmed the 
many previous studies that suggested a higher X-ray AGN population in clusters 
from surface density arguments alone \citep[e.g.][]{cappi01,sun02,ruderman05}, 
yet it is still not clear if the X-ray AGN fraction is higher than the field 
value. To date there is only weak evidence that the X-ray AGN fraction in 
clusters is comparable to the fraction in field early-type galaxies 
\citep{lehmer07,sivakoff08,arnold09}. 
One of the virtues of the emission-line galaxy studies as a function of 
environment is that they can directly calculate the fraction of a given 
galaxy population that hosts AGN as a function of environment, even though 
this technique appears to systematically miss AGN in the densest regions 
relative to X-ray and radio selection. 

In addition to a local comparison between AGN in different environments, 
measurement of the evolution of the AGN population in clusters can 
constrain the formation epoch for their supermassive black holes and the 
extent of their coevolution with the cluster galaxy population. The 
key early work on the evolution of galaxies in clusters was by 
\citet{butcher78,butcher84}, who observed a substantial increase in the 
fraction of blue galaxies in higher-redshift clusters. The Butcher-Oemler
effect is interpreted as an increase in the amount of star formation 
and has been confirmed by many other indicators, in particular \oii\ 
emission-line galaxy fractions \citep{poggianti06} and an increase in the 
number of $24\um$ sources in \spitzer\ observations 
of distant clusters \citep{bai07,saintonge08}. The observed increase brings the 
star formation rate (SFR) in cluster galaxies closer to those in the field. 
At a redshift of $z\sim1$ and higher, observations with \spitzer\ even find 
that galaxies in denser environments have higher star formation rates than 
lower-density regions \citep{elbaz07}, which is opposite the trend observed
in the local universe. Similar results have also been 
found with deep UV data \citep{heinis07}. The situation is less clear when 
star formation is measured with the \oii\ emission line because while 
\citet{poggianti08} find that star formation does not strongly depend on 
environment, \citet{cooper08} find the specific star formation rate has a 
similar dependence on environment at $z=0$ and $z=1$, although the total star 
formation rate is higher in clusters at $z=1$ than in the field. 

The existence of the Butcher-Oemler effect and the many indirect arguments 
outlined above for a connection between star formation and black hole 
accretion suggest that there should be an increased AGN population in 
high-redshift clusters. An early study of the high-redshift cluster 3C295 at 
$z=0.46$ by \citet{dressler83} found evidence for three AGN and was an 
indication that this may be the case; however, their relative scarcity 
precluded a detailed statistical study or targeted studies to deliberately 
identify cluster AGN. This situation changed dramatically 
with the launch of \chandra, whose superb sensitivity and angular resolution 
produced a dramatic increase in efficiency for searches for AGN, particularly 
lower-luminosity sources. Just as the case for local clusters, \chandra\ 
observations of distant clusters have revealed substantial populations of 
point sources \citep{cappelluti05,gilmour09}. Spectroscopic confirmation 
that these point sources are associated with cluster members has been 
more challenging \citep{johnson03,demarco05}, but in \citet{eastman07} we 
combined new observations of MS2053.7-0449 ($z=0.58$) with archival data 
on three additional, $z>0.5$ clusters and found an 
approximately order of magnitude increase in the fraction of $M_R < -20$ mag 
galaxies that hosted AGN more luminous than $L_{X,H} \geq 10^{43}$ \ergs\ in 
the hard X-ray band (2--10 keV) relative to the sample of ten low-redshift 
$z < 0.32$ clusters in \citet{martini07}. These results have since been 
strengthened with detailed studies of clusters at $z\sim1$ with XMM 
\citep{vanbreukelen09} and measurements of surface density excesses in 
clusters to $z\sim 1.5$ \citep{galametz09}. 

In addition to their application to the coevolution of black holes and 
galaxies, an increase in the AGN fraction in clusters may also impact the 
ICM. At low redshifts many studies have shown that 
AGN feedback is a viable explanation for the absence of substantial reservoirs
of cold gas at the centers of clusters \citep[for a recent review 
see][]{mcnamara07}. This feedback is ascribed to AGN associated with 
the central cluster galaxy, which is almost invariably a luminous radio 
source. In our studies of X-ray AGN this is almost the only cluster 
galaxy in which we are {\it insensitive} to the presence of an AGN because 
it is challenging to measure even a bright nuclear point source when 
juxtaposed with the extended emission from the ICM that often peaks near the
central cluster galaxy. Nevertheless, the
evolution of AGN in other cluster galaxies is likely to be connected to the 
evolution of the central AGN as the stars in the most luminous cluster galaxies 
have comparable ages. An increase in the net energy production by AGN in 
higher-redshift clusters is of interest because energy input during 
cluster formation has been invoked as an explanation for the minimum entropy 
level in the ICM \citep{kaiser91,evrard91}. AGN remain perhaps the most 
viable mechanism, if only because most others can be ruled out 
\citep{kravtsov04,conroy08}, although the details of how AGN feedback couples 
to the ICM remain uncertain. Outside of the central galaxy, an increase 
in the number of other AGN associated with clusters of galaxies may also 
affect measurement of other cluster properties 
\citep{branchesi07b,bignamini08}. Finally, an analogous increase in the 
radio-loud AGN population in high-redshift clusters may contaminate searches 
for clusters via the Sunyaev-Zel'dovich effect \citep{sunyaev70} at mm and 
cm wavelengths. As many searches for clusters that exploit this effect are in 
progress, it is important to characterize the potential impact of 
evolution of the cluster AGN population on these experiments \citep[e.g.][]{lin07}. 

In the next section we describe our expanded high-redshift data, as well as 
the selection criteria for X-ray AGN we employ at all redshifts. We then 
describe our new observations of low-redshift clusters in \S\ref{sec:lowzdata}. 
These two datasets are combined to calculate the cluster AGN fraction and 
its evolution in \S\ref{sec:fa}, followed by an examination of the properties 
of the cluster AGN in \S\ref{sec:agn}. We discuss the implications of 
these results, particularly on the coevolution of black holes and galaxies, 
in \S\ref{sec:dis} and conclude with a summary of our results. 
Throughout this paper we assume that the cosmological parameters are: 
($\Omega_M, \Omega_\Lambda, h$) = (0.3, 0.7, 0.7) where $H_0 = 100h$ \kms 
Mpc$^{-1}$. All absolute magnitudes quoted in this paper assume $h = 0.7$. 

\section{High-Redshift Data} \label{sec:highzdata} 


\begin{deluxetable*}{lrrlrccccl}
\tablecolumns{10}
\tablewidth{7.0truein}
\tabletypesize{\scriptsize} 
\tablecaption{High-Redshift Cluster Sample\label{tbl:highz}}
\tablehead{
\colhead{Cluster} &
\colhead{$\alpha_c$} &
\colhead{$\delta_c$} &
\colhead{$z$} & 
\colhead{$\sigma$ [km/s]} & 
\colhead{$\sigma$ Ref} & 
\colhead{$T_X$ [keV]} & 
\colhead{$T_X$ Ref} & 
\colhead{$R_{200}$ [Mpc]} & 
\colhead{Spectra} \\ 
\colhead{(1)} &
\colhead{(2)} &
\colhead{(3)} &
\colhead{(4)} & 
\colhead{(5)} & 
\colhead{(6)} & 
\colhead{(7)} & 
\colhead{(8)} & 
\colhead{(9)} & 
\colhead{(10)} 
}
\startdata
MS 1621.5+2640    & 16:23:34.9 & +26:34:21 & 0.43  &  735 & 1 & 7.6 & 1 & 1.42 & SESXI \\ 
3C 295            & 14:11:20.5 & +52:12:09 & 0.46  & 1642 & 1 & 5.3 & 1 & 3.12 & SESXI \\
MS 0451.6-0305    &  4:54:11.1 & -03:00:55 & 0.538 & 1371 & 2 & 8.1 & 1 & 2.49 & ChaMP \\
MS 0015.9+1609    &  0:18:33.5 & +16:26:06 & 0.541 & 1234 & 2 & 9.4 & 2 & 2.24 & ChaMP \\
RX J0848.7+4456   &  8:48:47.6 & +44:56:16 & 0.574 &  670 & 3 & 3.2 & 3 & 1.19 & SESXI \\
MS 2053.7-0449    & 20:56:21.3 & -04:37:49 & 0.583 &  865 & 4 & 5.2 & 1 & 1.53 & SESXI,ChaMP \\
RX J0542.8-4100   &  5:42:49.8 & -41:00:07 & 0.634 & 1101 & 3 & 7.9 & 3 & 1.89 & ChaMP \\
RX J2302.8+0844   & 23:02:48.3 & +08:43:48 & 0.722 &  993 & 3 & 6.6 & 3 & 1.61 & ChaMP \\
MS 1137.5+6625    & 11:40:22.1 & +66:08:14 & 0.782 &  967 & 3 & 6.3 & 1 & 1.52 & ChaMP \\
RX J1317.4+2911   & 13:17:22.0 & +29:11:24 & 0.805 &  531 & 3 & 2.2 & 1 & 0.82 & SESXI \\
RX J1716.4+6708   & 17:16:49.3 & +67:08:25 & 0.813 & 1445 & 1 & 6.6 & 1 & 2.22 & SESXI,ChaMP \\
MS 1054-03        & 10:56:55.7 & -03:37:39 & 0.831 & 1156 & 5 & 7.8 & 1 & 1.76 & ChaMP \\
RDCS J0910+5422   &  9:10:44.7 & +54:22:04 & 1.11  &  675 & 6 & 3.5 & 1 & 0.87 & SESXI \\
Lynx E            &  8:48:58.3 & +44:51:51 & 1.261 &  740 & 7 & 3.8 & 4 & 0.88 & SESXI \\
Lynx W            &  8:48:34.2 & +44:53:35 & 1.27  &  650 & 8 & 1.7 & 4 & 0.77 & SESXI 
\enddata
\tablecomments{
Cluster sample and properties derived from the present study. Columns are: (1) 
Cluster name; (2 and 3) RA and DEC for the centroid of the extended 
X-ray emission; (4) redshift; (5) velocity dispersion; (6) reference for the velocity dispersion; (7) X-ray temperature in keV; (8) reference for the X-ray 
temperature; (9) estimate of the virial radius in Mpc \citep[e.g.,][]{treu03}; (10) 
origin of most of the spectra. 
References for velocity dispersion are: 1: \citet{girardi01}; 2: \citet{carlberg96}; 3: derived from the X-ray temperature following \citet{xue00}; 4: \citet{tran05}; 5: \citet{tran07}; 6: \citet{mei06}; 7: from weak lensing 
estimate \citet{jee06}; 8: \citet{stanford01}. 
References for X-ray temperatures are: 1: \citet{vikhlinin02}; 2: \citet{ebeling07}; 3: \citet{ettori04}; 4: \citet{jee06}. 
}
\end{deluxetable*}

Two large surveys have obtained redshifts for substantial numbers of galaxies 
with X-ray counterparts in many deep, archival \chandra\ observations that 
include substantial numbers of high-redshift clusters of galaxies. These are 
the Serendipitous Extragalactic X-ray Source Identification Program 
\citep[SEXSI;][]{harrison03,eckart05,eckart06} and the Chandra Multiwavelength 
Project \citep[ChaMP;][]{kim04a,kim04b,green04,silverman05a}. 
We have investigated the fields surveyed by both SEXSI and ChaMP to identify 
datasets that contain clusters of galaxies with $z>0.4$ and have sufficient 
depth to identify $L_{X,H} \geq 10^{43}$ \ergs\ (rest frame 2--10 keV) AGN 
at the cluster redshift. 

The SEXSI survey published spectroscopic redshifts for 27 archival 
\chandra\ observations in \citet{eckart06} that were selected to 
identify hard X-ray sources over the flux range of $10^{-13} - 10^{-15}$ 
\ergs cm$^{-2}$ and isolate those responsible for the hard X-ray background. 
The specific selection criteria for the fields were that they must be 
high Galactic latitude ($|b|>20^{\circ}$) and be obtained with either the 
I or S modes of the Advanced Camera for Imaging Spectroscopy 
\citep[ACIS;][]{bautz98} when no grating was used. The X-ray luminosities 
quoted by SEXSI are based on 
spectral fits that assume a $\Gamma = 1.5$ power law and intrinsic absorption 
$N_H$ at the source redshift, although they quote the observed luminosities 
(not corrected for obscuration) and provide the best-fit $N_H$ value. 
The average spectroscopic completeness is 67\% (see \S\ref{sec:completeness} 
below) for sources with $R < 24.4$ mag on the Vega system. 
Nine of the 27 SEXSI fields include clusters of galaxies 
with $z > 0.4$ and we include seven\footnote{RX J1350.0+6007 was not targeted 
for spectroscopy and the X-ray data for CL0442+0202 ($z=1.11$) were 
sufficiently shallow ($t = 44$ks) that 
they may not be complete to $L_{X,H} = 10^{43}$ \ergs. In addition, 
\citet{stern03} classify CL0442+0202 as an overdensity that has not yet 
collapsed, rather than as a cluster.}  
in our sample. As one field contains 3 clusters, we list nine clusters from 
SEXSI in Table~\ref{tbl:highz}.

The ChaMP survey published spectroscopic redshifts for 20 archival 
\chandra\ observations in \citet{silverman05a} that were similarly selected for 
depth, high Galactic latitude ($|b|>20^{\circ}$), and no special observing 
modes. The spectroscopic completeness of ChaMP is 77\% at $r'<22.5$ mag, 
where $r'$ is on the SDSS photometric system \citep[][and $r'_{AB} = 
R_{Vega} + 0.17$]{fukugita96}. 
Their X-ray luminosities are based on spectral fits 
that assume a $\Gamma = 1.9$ power law and intrinsic absorption $N_H$ at the 
source redshift, as well as the appropriate Galactic absorption, although 
they also quote the observed luminosities (only corrected for Galactic 
absorption). 
The final sample presented in \citet{silverman05a} was restricted to X-ray 
sources with $L_X > 10^{42}$ keV in the 2--8 keV band in order to insure all 
are AGN. Most (69\%) are spectroscopically classified as broad-line
AGN (BLAGN). Nine of these 20 ChaMP fields include clusters of galaxies with 
$z > 0.4$ and we include 
eight\footnote{We exclude CL J0152.7-1357 ($z=0.831$) because the exposure time 
is shorter than the others at $t = 34.6$ ks and therefore the X-ray data may 
not be complete to $L_{X,H} = 10^{43}$ \ergs.} of these in our study (see 
Table~\ref{tbl:highz}). Two of these clusters are common to both ChaMP and 
SEXSI (MS2053.7-0449 and RXJ1716.4+6708) and therefore the final sample 
has fifteen clusters with $z > 0.4$. While spectroscopic data for X-ray 
sources in other high-redshift clusters exist \citep[e.g.][]{johnson06}, we 
limit our high-redshift sample to these fifteen to maximize the uniformity of 
the dataset. 


\begin{deluxetable*}{lllllllll}
\tablecolumns{9}
\tablewidth{7.0truein}
\tabletypesize{\scriptsize}
\tablecaption{High-Redshift Cluster AGN Sample\label{tbl:highzagn}}
\tablehead{
\colhead{AGN} &
\colhead{Cluster} & 
\colhead{$z$} &
\colhead{$R$ [mag]} &
\colhead{log $L_{X,H}$ [\ergs]} & 
\colhead{$\delta v/\sigma$} &
\colhead{$\Delta R$ [arcmin]} &
\colhead{$R/R_{200}$} &
\colhead{Class} \\
\colhead{(1)} &
\colhead{(2)} &
\colhead{(3)} &
\colhead{(4)} &
\colhead{(5)} &
\colhead{(6)} &
\colhead{(7)} &
\colhead{(8)} & 
\colhead{(9)} 
}
\startdata
CXOSEXSI J141127.4+521131&	3C295		&0.451	&19.78	&43.4	&1.13	&1.23	&0.14	&ALG \\
CXOSEXSI J141123.4+521331&	3C295		&0.472	&19.05	&43.8	&1.5	&1.45	&0.16	&BLAGN \\
E0015+162		&	MS0015.9+1609	&0.553	&18.41	&45.48	&1.89	&3.35	&0.58	&BLAGN \\
CXOSEXSI J084858.0+445434&	RX J0848.7+4456 &0.573	&19.58	&43.8	&0.28	&2.5	&0.83	&BLAGN \\
CXOMP J054248.2-410140	&	RDCSJ0542-4100	&0.634	&20.64	&43.24	&0	&1.58	&0.32	&NELG \\
CXOMP J054251.4-410205	&	RDCSJ0542-4100	&0.637	&19.63	&43.35	&0.5	&1.99	&0.33	&ALG \\
CXOMP J054259.5-410241	&	RDCSJ0542-4100	&0.638	&20.50	&43.37	&0.67	&3.16	&0.63	&NELG \\
CXOMP J054240.8-405626	&	RDCSJ0542-4100	&0.639	&20.89	&43.67	&0.83	&4.05	&0.81	&NELG \\
CXOMP J054255.0-405922 	&	RDCSJ0542-4100	&0.644	&22.08	&43.08	&1.67	&1.24	&0.25	&NELG \\
CXOMP J114022.0+660816	&	MS1137+6625	&0.786	&20.37	&43.24	&0.7	&0.04	&0.01	&BLAGN \\
CXOSEXSI J171636.9+670829&	RXJ1716.4+6708	&0.795	&22	&44	&2.06	&1.19	&0.24	&ELG \\
CXOSEXSI J131718.8+291111&	RX J1317.4+2911	&0.803	&21.98	&43.3	&0.63	&0.68	&0.38	&BLAGN \\
CXOSEXSI J171703.8+670900&	RXJ1716.4+6708	&0.812	&21.79	&43	&0.11	&1.53	&0.31	&ELG \\
CXOSEXSI J171714.5+671136&	RXJ1716.4+6708	&0.815	&22.68	&43.2	&0.23	&4.02	&0.82	&ELG \\
CXOMP J105650.6-033508	&	MS 1054-03	&0.818	&21.76	&43.22	&1.84	&2.82	&0.73	&BLAGN \\
CXOU J091043.3+542152	&	RDCSJ0910+5422	&1.104	&24  	&43.06	&1.26	&0.29	&0.16	&AGN2 \\
CXOSEXSI J084905.3+445203&	LynxE		&1.266	&24.61	&43.8	&1.11	&1.27	&0.74	&ELG \\
CXOSEXSI J084831.6+445442&	LynxW		&1.267	&25.42	&43.2	&0.61	&1.23	&0.8	&ELG 
\enddata
\tablecomments{
AGN in high-redshift clusters of galaxies. Columns are: (1) AGN name; (2) 
Cluster; (3) AGN redshift; (4) $R-$band magnitude; (5) Rest-frame, hard-X-ray 
luminosity (2--10 keV); (6) Velocity offset from the cluster systemic velocity 
normalized by the cluster velocity dispersion; (7) Projected radial offset 
relative to the centroid of the X-ray gas in arcminutes; (8) Projected radial 
offset normalized by the cluster virial radius; (9) Spectroscopic 
classification. The $R-$band magnitude of E0015+162 is from \citet{orndahl03}. 
The remaining values are from either \citet{eckart06} for the SEXSI sample or 
from \citet{silverman05a} for the ChaMP sample (although corrected from $r'$ 
to $R$ as noted in Section~\ref{sec:highzdata}). The 2--8 keV X-ray 
luminosities from \citet{silverman05a} have been corrected to the 2--10 keV 
band as described in Section~\ref{sec:highzdata}. 
}
\end{deluxetable*} 

We have also compiled additional data for each cluster listed in 
Table~\ref{tbl:highz} that will be important for our subsequent 
analysis. One quantity is the center of the cluster, which is 
needed to determine if a given AGN falls within the projected 
virial radius of the cluster. We associate the center of each cluster 
with the centroid of the extended X-ray emission. While these coordinates
do not always agree with the standard coordinates quoted in the literature, 
this assumption makes our analysis more uniform. The redshift and velocity 
dispersion are also needed to determine if an AGN is within the cluster.
In most cases velocity dispersions for these clusters are available 
in the literature and we quote the origin of the measurement we adopt in the 
table. When the velocity dispersion has not been measured, we estimate 
this quantity from the X-ray temperature and the $\sigma - T_X$ relationship 
from \citet{xue00}. Specifically, we used the relation 
$\sigma = 10^{2.51\pm0.01} T^{0.61\pm0.01}$ \kms\ derived from their combined 
group and cluster sample with orthogonal distance regression 
\citep{feigelson92}. Based on their data, we estimate that there is 
a 30\% uncertainty in $\sigma$ at fixed $T$. 

One potential concern for our subsequent analysis is that the 
\citet{xue00} $\sigma-T$ relation may not hold at higher redshift. 
\citet{lubin04} investigated this point for several optically-selected 
clusters and found that they were 2-9 times cooler than expected from 
the local relation; however, the difference was much less stark for 
X-ray selected, high-redshift clusters similar (and in several cases identical 
to) those presented here. \citet{fang07} showed that high-redshift, 
X-ray selected clusters are consistent with the low-redshift $L_X - \sigma$ 
relation, although spectroscopically-selected groups and clusters do not 
agree as well \citep[see also][]{andreon08}. 

Finally, we have calculated the projected size of the virial radius for 
each cluster following \citet{treu03} and throughout this paper we associate 
the virial radius with $R_{200}$, the radius within which the cluster is a 
factor of 200 overdensity. Of the three clusters we have in common with 
\citet{poggianti06}, for 3C~295 and MS1054-03 we adopt nearly the same 
$\sigma$ and our $R_{200}$ estimate is nearly identical to theirs, 
while for MS0015.9+1609 we adopt a slightly larger velocity 
dispersion (1234 \kms\ from \citet{carlberg96} rather than their 984 \kms) and 
consequently infer a larger radius. 

Because the most relevant ChaMP measurements are the 2--8 keV luminosity, 
rather than 2--10 keV luminosity, we multiply the ChaMP 2--8 keV luminosities 
by a factor of 1.2. This correction factor was calculated for a $\Gamma=1.7$ 
power law with PIMMS. There is some uncertainty in this correction factor 
because not all AGN have this power-law form, particularly as we assume this 
correction for their observed rather than intrinsic (unobscured) spectra, but 
this is not a significant effect compared to other sources of systematic errors 
that we discuss below. There are no additional AGN from ChaMP that enter the 
sample after this step because there are none just below the $10^{43}$ \ergs\ 
threshold in the 2--8 keV band. We also estimated the difference in luminosity 
for an AGN calculated with the $\Gamma=1.5$ power law employed by SEXSI, the 
$\Gamma=1.9$ employed by ChaMP, and a $\Gamma=1.7$ power law to determine 
if these differences would cause any sources to fall in or out of the same 
and none would do so. 
In the two clusters observed by both ChaMP and SEXSI, there is one 
cluster AGN common to both surveys: CXOSEXSI J171636.9+670829. The 
redshifts from the two surveys agree exactly ($z=0.795$) and the luminosities 
agree well: $L_{X,2-10} = 10^{44}$ \ergs and $L_{X,2-8} = 10^{43.88}$ \ergs. 

We also correct the ChaMP $r'$ measurements to the Vega $R$ band as discussed 
above. Based on the magnitudes of these sources and a simple $k-$correction, 
we estimate that none of these sources falls below our galaxy luminosity 
threshold. As these are fairly luminous AGN, in some cases the AGN may dominate 
the total flux and we may have overestimated the host galaxy luminosity. 
E0015+162 \citep{margon83} is the most X-ray luminous AGN in our sample by 
over an order of magnitude and is a useful case study to test the importance 
of this concern. 
This AGN has a total $R =18.41$ mag and a fainter host galaxy magnitude of 
$R = 19.8$ mag \citep{orndahl03}, which corresponds to a factor of 3.6 in 
flux. If the other AGN have similar or smaller $L_R/L_X$ ratios (such as 
due to obscuration), then we expect their AGN contribution to the measured 
$R-$band flux to be negligible because they are all much less luminous than 
E0015+162.

We identify AGN in these clusters with the following four criteria: 1) The hard 
X-ray luminosity must be $L_{X,H} \geq 10^{43}$ \ergs; 2) The AGN redshift 
must fall within $3\sigma$ of the cluster mean redshift, where $\sigma$ is the 
cluster velocity dispersion; 3) The AGN must fall within the projected virial 
radius $R_{200}$ of the cluster; 4) The absolute magnitude of the host galaxy 
must be greater than $M_R = M_R^*(z)+1$ mag. 
Most of these criteria were adopted from \citet{eastman07}, although the 
absolute magnitude criterion is different and we discuss our motivation 
for this choice in \S\ref{sec:richness} below. With these criteria we identify 
18 AGN in the 15 clusters with $z>0.4$, or an average of more than one per 
cluster. The properties of the $z>0.4$ AGN are presented in 
Table~\ref{tbl:highzagn}. 

\section{New Low-Redshift Observations} \label{sec:lowzdata} 


\begin{deluxetable*}{lcccrcccc}
\tablecolumns{9}
\tablewidth{7.0truein}
\tabletypesize{\scriptsize} 
\tablecaption{New Low-Redshift Clusters\label{tbl:lowz}}
\tablehead{
\colhead{Cluster} &
\colhead{$\alpha_c$} &
\colhead{$\delta_c$} &
\colhead{$z$} & 
\colhead{$\sigma$ [km/s]} & 
\colhead{$\sigma$ Ref} & 
\colhead{$T_X$ [keV]} & 
\colhead{$T_X$ Ref} & 
\colhead{$R_{200}$ [Mpc]} \\
\colhead{(1)} &
\colhead{(2)} &
\colhead{(3)} &
\colhead{(4)} &
\colhead{(5)} &
\colhead{(6)} & 
\colhead{(7)} & 
\colhead{(8)} & 
\colhead{(9)} 
}
\startdata
Abell 1240	& 11:23:37.3&	+43:06:54& 0.1590& 698	& 1 &  ... & ...  & 1.64  \\ 
Abell 1942	& 14:38:22.0&	+03:40:07& 0.2240& 903	& 2 &  5.6 & 1 & 1.96 \\ 
Abell 2125	& 15:41:13.2&	+66:16:01& 0.2465& 1113	& 3 &  3.2 & 2 & 2.39 \\ 
MS1455.0+2232	& 14:57:15.1&	+22:20:29& 0.2578& 1032	& 4 &  5.5 & 3 & 2.20 \\ 
ZwCl 1358.1+6245& 13:59:50.6&	+62:31:04& 0.3280& 1003	& 4 &  6.5 & 3 & 2.06 \\ 
MS1512.4+3647	& 15:14:22.4&	+36:36:21& 0.3720& 575	& 4 &  3.6 & 3 & 1.15 
\enddata
\tablecomments{
New low-redshift clusters and their properties derived from the present study. 
Columns are: (1) 
Cluster name; (2 and 3) RA and DEC for the centroid of the extended 
X-ray emission; (4) redshift; (5) velocity dispersion; (6) reference for the velocity dispersion; (7) X-ray temperature in keV; (8) reference for the X-ray 
temperature; (9) estimate of the virial radius in Mpc \citep{treu03}.
References for velocity dispersion are: 1: derived from the X-ray luminosity 
following \citet{xue00}; 2: derived from the X-ray temperature following 
\citet{xue00}; 3: \citet{miller04}; 4: \citet{borgani99}.
References for X-ray temperatures are: 1: \citet{ota04}; 2: \citet{wang04}; 3: \citet{mushotzky97}. 
}
\end{deluxetable*}

AGN more luminous than $L_{X,H} = 10^{43}$ \ergs\ are sufficiently rare in 
low-redshift clusters that Poisson uncertainties (as opposed to sources of 
systematic errors) from the low-redshift sample may dominate the statistical 
significance of any evidence of evolution. Our previous study of ten 
clusters with $z<0.32$ only identified one AGN above this luminosity threshold 
\citep{martini06}, while our more recent observations of three additional 
clusters (all at $z<0.08$) have identified only one additional AGN above this 
luminosity \citep{sivakoff08}. We have therefore studied six additional 
clusters with $0.15 < z < 0.4$ to find other X-ray AGN more luminous than 
$L_{X,H} = 10^{43}$ \ergs\ with a combination of \chandra\ archival data and 
follow-up spectroscopy of candidate cluster X-ray AGN at the MDM Observatory. 
These clusters were selected to be the nearest massive clusters in the 
\chandra\ archive whose estimated virial radii fit within the \chandra\ ACIS 
field of view (FOV) and were accessible during our observing runs. The new 
clusters and their physical properties are listed in Table~\ref{tbl:lowz}. 

\subsection{\chandra\ X-ray Analysis}


\begin{deluxetable}{lrrcc}
\tabletypesize{\footnotesize}
\tablewidth{0pt}
\tablecaption{{\it Chandra} Observation Logs\label{tab:xobs}}
\tablehead{
\colhead{Cluster} &
\colhead{OBSID} &
\colhead{Detector} &
\colhead{T} &
\colhead{$L_{X,H,{\rm Lim}}$} \\
& & & (ks) & ($10^{41} {\rm \, erg \, s}^{-1}$)\\
\colhead{(1)} &
\colhead{(2)} &
\colhead{(3)} &
\colhead{(4)} &
\colhead{(5)}
}
\startdata
Abell~1240       & \dataset[ADS/Sa.CXO#obs/04961]{4961} & ACIS-I  & 51.3 & $1.2$\\
Abell~1942       & \dataset[ADS/Sa.CXO#obs/03290]{3290} & ACIS-I  & 57.5 & $2.2$\\
Abell~2125       & \dataset[ADS/Sa.CXO#obs/02207]{2207} & ACIS-I  & 81.5 & $1.9$\\
MS~1455.0+2232   & \dataset[ADS/Sa.CXO#obs/04192]{4192} & ACIS-I  & 91.9 & $1.8$\\
ZwCl~1358.1+6245 & \dataset[ADS/Sa.CXO#obs/00516]{516}  & ACIS-S3 & 53.0 & $2.6$\\
MS~1512.4+3647   & \dataset[ADS/Sa.CXO#obs/00800]{800}  & ACIS-S3 & 36.4 & $4.6$
\enddata
\tablecomments{{\it Chandra} Observation Log. Columns are:
(1) Cluster targeted;
(2) Observation ID of {\it Chandra} data;
(3) Detector used;
(4) Usable exposure;
(5) Estimate of the $2.0$--$8.0 {\rm \, keV}$ luminosity limit of the observation for a cluster galaxy.}
\end{deluxetable}

The X-ray observations were processed following the same techniques employed by
\citet{sivakoff08}. We reduced all data using {\sc ciao 3.4}%
\footnote{\url{http://asc.harvard.edu/ciao/}.}
with {\sc caldb 3.3.0.1} and NASA's {\sc ftools 6.0}%
\footnote{\url{http://heasarc.gsfc.nasa.gov/docs/software/lheasoft/}%
\label{ftn:heasoft}.}. The observations are summarized in Table~\ref{tab:xobs}.
Only minor differences in reduction were required for these archival
observations. The majority of the clusters had data with an aimpoint centered on
the four ACIS-I chips ($\sim 17\arcmin$ FOV) and frame times of $3.1 {\rm \,
s}$. These data were telemetered and cleaned in Very Faint mode. The more
distant clusters, ZwCl~1358.1+6245 and MS~1512.4+3647, were observed with the
aimpoint placed on the ACIS-S3 detector ($8.4'$ FOV) and had frame times
of $3.3 {\rm \, s}$. Their data were telemetered and cleaned in Faint mode, and
thus have a slightly higher background. As all observations were operated at
$-120 ^{\circ} \,{\rm C}$ the X-ray data were corrected for the time dependence
of the gain and the charge-transfer inefficiency with their photon energies
determined using the gain file acisD2000-01-29gain$\_$ctiN0006.fits. The
archival data of all observations already had applied the newest tools to
detect hot pixels and cosmic ray afterglows. We only consider events with ASCA
grades of 0, 2, 3, 4, and 6. Known aspect offsets were applied
for each observation. All observations were corrected for quantum efficiency
degradation and had exposure maps determined at $1.5 {\rm \, keV}$. We excluded
bad pixels, bad columns, and columns adjacent to bad columns or chip node
boundaries. We also filtered out times when the blank-sky rate was more than 
three times
the expected blank-sky rate derived from calibrated blank-sky backgrounds to
avoid the most extreme periods of high background (``background flares'') that
\chandra\ may encounter. MS~1512.4+3647 had two separate pointings and 
this introduced difficulties into our standard processing. We therefore 
excluded the shorter second pointing, which accounted for less than
25\% of the total integration time. 

To detect X-ray sources that are potential X-ray AGN in these clusters, we
applied the wavelet detection algorithm ({\sc ciao wavdetect}) with
scales ranging from 1 to 64 pixels in steps of $\sqrt{2}$ factors and required 
a source detection threshold of $10^{-6}$. Source detection was only performed 
in regions with an exposure of greater than 10\% of the total for the 
observation.  Our source detection threshold corresponds to $\la 4$ falsely 
detected X-ray sources (due to a statistical fluctuation) for each observation. 
Using \citet{kim07}, we have estimated the statistical X-ray positional 
uncertainty (1$\sigma$) due to {\sc wavdetect}. In Table~\ref{tab:xobs}, we 
list an estimated limiting X-ray luminosity for each observation that 
corresponds to five counts on axis \citep[for consistency with][]{martini06}. 
For our analysis we concentrated on sources with at least 20 broad 
(0.3--8.0 keV) X-ray counts. These sources are unlikely to be due to 
statistical fluctuations except where they are coincident with ICM emission.

We used ACIS Extract 3.131\footnote{http://www.astro.psu.edu/xray/docs/TARA/ae\_users\_guide.html}
to create source extraction regions enclosing 90\% of the flux in the X-ray PSF
and to determine a masking radius that encircled 97\% of the flux. For most of
the sources, whose photons had median energies of $\sim 0.6$--$2.6 {\rm \,
keV}$, we determined the regions assuming the PSF at $1.497 {\rm \, keV}$. A few
sources had harder emission and their PSF was calculated assuming an energy of
$4.51 {\rm \, keV}$. In a relatively small number of crowded regions, the
PSF fraction was reduced to prevent overlapping source extraction regions. We
also used ACIS Extract to correct the ({\sc ciao wavdetect}) position to the
mean position of detected events for sources within 5\arcmin of the observation
aimpoint or to the position that best correlated with the PSF for sources
beyond 5\arcmin of the observation aimpoint. These new positions were registered
with an optical catalog from $R-$band images (see below) to correct the absolute
astrometry and determine the absolute astrometric precision of each \chandra\ 
observation (0.3--0.5\arcsec). The statistical significance of each detection
was added in quadrature with the absolute astrometric precision to estimate the
total X-ray positional precision.
We measured the counts in three energy ranges: the broad (0.3--8 keV), soft 
(0.3--2 keV), and hard (2.0--8.0 keV) bands. The observed
fluxes in these bands were derived assuming a $\Gamma=1.7$ power-law spectrum
with Galactic absorption. We then calculated the rest-frame luminosity in 
the broad band (0.3--8 keV) and the classic hard band (2--10 keV) for all 
sources with redshifts (see \S\ref{sec:spectra}). 

\subsection{MDM Photometry} \label{sec:phot} 

$R-$band images of these clusters were obtained at the MDM Observatory 2.4m
Hiltner telescope with the Echelle CCD camera during a run from the night of 28
May 2007 to 3 June 2007. Because the FOV of the CCD camera
($\sim 9.5' \times 9.5'$) is smaller than the ACIS-I FOV ($\sim 17' \times
17'$), we imaged a $2 \times 2$ mosaic to cover the \chandra\ area, with each
panel consisting of $3\times300 {\rm \, s}$ exposures. All images were trimmed,
bias-subtracted and flat-fielded with the {\sc ccdproc} package within
IRAF\footnote{IRAF is distributed by the National Optical Astronomy Observatory,
which is operated by the Association of Universities for Research in Astronomy
(AURA) under cooperative agreement with the National Science Foundation.}.

Sources were cataloged with the SExtractor package \citep{bertin96}.
Aperture magnitudes from these catalogs were calibrated with multiple 
observations of standard star fields from the data compiled by P. B. 
Stetson\footnote{\url http://cadcwww.hia.nrc.ca/standards}
onto the Vega magnitude system. Only data from the last night, which includes
each quadrant of Abell~1240 and ZwCl~1358.1+6245, the north-east quadrant of
MS~1512.4+3647, and $1\times300 {\rm \, s}$ exposures of each quadrant of
Abell~2125, were taken under photometric conditions. Our derived photometric
solution for this night was precise to 0.03 mag. As all of these clusters 
except for Abell~2125 were imaged with SDSS, we cross-correlated aperture 
magnitudes from all images on this run with stars in the SDSS DR5 catalog. 
After correcting to R 
(Vega)\footnote{\url http://www.sdss.org/dr7/algorithms/sdssUBVRITransform.html\#Lupton2005},
our derived photometric solution for 3 June, which includes a color correction
term, is accurate to 0.01 mag and precise to 0.06 mag. The poorer precision
compared to our photometric solution appears to be only partially due to the
dispersion in the Vega correction (overlapping sources between quadrants of 
our own observations indicate typical
photometric precisions of $0.05-0.08 {\rm \, mag}$). We therefore adopted the
SDSS cross-calibration technique to photometrically correct all observations on
non-photometric nights, except for observations of Abell~2125. For Abell~2125,
non-photometric observations were cross calibrated with the single photometric
exposures for Abell~2125. As we do not have complete multi-band data, we report
only the magnitudes assuming no color correction. The exclusion of the color
correction term does not significantly decrease the precision of our photometric
solutions.

We calculated astrometric solutions for the images with the WCSTools package
\citep{mink02}, package and then produced the final, calibrated mosaics with the
SWARP\footnote{http://terapix.iap.fr/rubrique.php?id\_rubrique=49} package. A
final source catalog was extracted with SExtractor and used to register the
astrometry of the X-ray observations. We consider only the SExtractor AUTO
magnitudes, which is an automatic aperture magnitude designed to give precise
estimates of total magnitudes for galaxies. As nearby, detected neighbors are
removed and replaced by mirroring the opposite side of the aperture where
available, these magnitudes are suitable for our relatively crowded fields. All
X-ray sources that would be more luminous than $L_{X,H} = 10^{43}$ \ergs\ at the
cluster redshift that were also associated with galaxies and that would be more
luminous than $M_R^*(z)+1$ at the cluster redshift were then targeted for
the highest-priority spectroscopic observations, with the exception of sources 
heavily contaminated by ICM emission. We also identified other 
candidate cluster X-ray AGN, specifically those that would have $L_{X,H} \geq 10^{42}$ 
\ergs, as lower-priority spectroscopic targets. 

\subsection{MDM Spectroscopy} \label{sec:spectra} 

We obtained low-resolution spectroscopy of these candidates with the 2.4m
Hiltner telescope with the CCDS, a Boller \& Chivens spectrograph, during a run
from the night of 28 April 2008 to 3 May 2008. The slit widths were 
determined by the nightly seeing conditions and were either $1.0''$ or 
$1.5''$. At least two exposures of every candidate were obtained and total 
exposure times varied from $120 {\rm \, s}$ to $9000 {\rm \, s}$. Five sets of 
internal and twilight flats were taken over the entire run, while comparison 
lamps were observed before and/or after every candidate.

The files were trimmed and bias-subtracted with the {\tt ccdproc} package 
within IRAF and bad pixels were determined from a ratio of flat-field
images and were fixed in every image. The individual flat-field images from
internal lamps revealed a complex wavelength and slit-dependent flat-field, most
likely due to some reflection. To model this complex response, we first median
smoothed the internal flat-fields (over $11\times11$ pixels) and then
Gaussian-smoothed ($\sigma=11$ pixels) over the dispersion axis. The ratio of
the internal flat-field to the modeled internal flat-field was adopted as the
true internal flat-field. An illumination correction was then created from the
twilight flat-fields and applied to make the final set of flat-field corrections
to remove fringing in the spectra. After each spectrum was properly
flat-fielded, we rejected cosmic rays using L.A. Cosmic\footnote{\url
http://www.astro.yale.edu/dokkum/lacosmic/} \citep{vandokkum01}. A fourth 
order wavelength-solution
was calculated for each set of HgNe comparison spectra, resulting in a typical
RMS of $\sim 0.1$\AA\ pixel$^{-1}$. Thereafter, standard aperture 
extraction of the spectra was used to remove the night sky emission and produce
one-dimensional, logarithmically interpolated spectra with a dispersion of
$\sim$ 3\AA\ pixel$^{-1}$. The spectra extend from approximately 3650\AA\ to
7250\AA. We extracted both the signal and noise for each final spectrum of a 
source.


\begin{deluxetable*}{llcccrrrrrc}
\tablecolumns{11}
\tabletypesize{\footnotesize}
\tablewidth{7.0truein}
\tablecaption{New, Lower-Luminosity Cluster X-ray AGN \label{tbl:xcagn}}
\tablehead{
\colhead{CXOU ID} &
\colhead{$z$} &
\colhead{$z$ ref} &
\colhead{$R$} &
\colhead{$R$ flag} &
\colhead{$f_{X,S}$} &
\colhead{$f_{X,H}$} &
\colhead{$f_{X,B}$} &
\colhead{$L_{X,B}$} &
\colhead{$L_{X,H}$} &
\colhead{X flag} \\
\colhead{(1)} &
\colhead{(2)} &
\colhead{(3)} &
\colhead{(4)} &
\colhead{(5)} &
\colhead{(6)} &
\colhead{(7)} &
\colhead{(8)} &
\colhead{(9)} &
\colhead{(10)} &
\colhead{(11)}
}
\startdata
J135950.5+623106.3 & $0.32717\pm0.00038$ & 1 & $17.80\pm0.05$ & 3 & $9.5^{+1.1}_{-1.0}$    & $8.7^{+2.8}_{-2.4}$    & $20.6^{+2.2}_{-2.1}$  & $70.6^{+7.6}_{-7.2}$ & $46.2^{+5.0}_{-4.7}$  & 1 \\
J143821.8+034013.3 & 0.22479             & 2 & $16.44\pm0.06$ & 3 & $2.20^{+0.91}_{-0.76}$ & $2.3^{+1.9}_{-1.4}$    & $4.8^{+1.7}_{-1.5}$   & $7.2^{+2.5}_{-2.2}$  & $4.7^{+1.7}_{-1.4}$   & 1 \\
J145714.7+221933.6 & $0.24852\pm0.00025$ & 1 & $20.04\pm0.07$ & 0 & $2.40^{+0.63}_{-0.53}$ & $3.7^{+1.3}_{-1.1}$    & $5.9^{+1.2}_{-1.0}$   & $11.2^{+2.2}_{-2.0}$ & $7.3^{+1.5}_{-1.3}$   & 0 \\
J145715.0+222034.5 & $0.25772\pm0.00015$ & 1 & $16.82\pm0.07$ & 0 & $20.2^{+4.9}_{-4.8}$   & $21.6^{+8.5}_{-8.3}$   & $44.4^{+9.1}_{-9.0}$ & $93^{+19}_{-19}$    & $61^{+12}_{-12}$ & 1 \\ 
J151422.5+363620.7 & 0.3718              & 3 & $18.05\pm0.06$ & 2 & $3.98^{+0.98}_{-0.89}$ & $3.2^{+2.5}_{-1.9}$    & $8.4^{+1.9}_{-1.8}$   & $38.1^{+8.8}_{-8.0}$ & $24.9^{+5.7}_{-5.2}$  & 1 \\
J154101.9+661627.1 & $0.24564\pm0.00045$ & 1 & $17.19\pm0.08$ & 2 & $2.78^{+0.62}_{-0.52}$ & $0.21^{+0.76}_{-0.41}$ & $4.63^{+1.0}_{-0.87}$ & $8.5^{+1.9}_{-1.6}$  & $5.5^{+1.2}_{-1.0}$   & 0 \\
J154101.9+661721.4 & 0.2567              & 4 & $19.36\pm0.08$ & 0 & $8.11^{+0.97}_{-0.88}$ & $7.1^{+1.7}_{-1.4}$    & $17.0^{+1.8}_{-1.6}$  & $34.3^{+3.6}_{-3.3}$ & $22.4^{+2.3}_{-2.1}$ & 0 \\
J154117.3+661923.6 & 0.2465              & 4 & $18.81\pm0.08$ & 0 & $2.08^{+0.58}_{-0.47}$ & $1.46^{+1.2}_{-0.82}$  & $4.15^{+1.1}_{-0.88}$ & $7.6^{+1.9}_{-1.6}$  & $5.0^{+1.3}_{-1.1}$ & 0 
\enddata
\tablecomments{{\it Chandra} Observation Log. Columns are:
(1) Name of X-ray source;
(2) Redshift
(3) References for redshift are: 1: this work; 2: SDSS \citep{adelman-mccarthy08}; 3 \citet{abraham98}; 4: \citet{miller04};
(4) $R-$band magnitude;
(5) Flags for photometry are: (0) no flag; (1) may be contaminated by nearby neighbors or bad pixels; (2) blended with nearby neighbors; (3) both;
(6--8) Soft [0.3--2 keV], Hard [2--8 keV], and Broad [0.3--8 keV] band flux 
in the observed frame in units of $10^{-15} {\rm \, erg \, s^{-1} cm^{-2}}$.
(9--10) Broad [0.3--8 keV] and Hard [2--10 keV] band luminosity in the 
rest-frame in units of $10^{41} {\rm \, erg \, s}^{-1}$ corrected for 
Galactic absorption.
(11) X-ray flags are: (0) no flag; (1) contaminated by ICM peak.
Note that CXOU J145715.0+222034.5 is the BCG and we subtracted a 
multi-component beta model for the ICM to compute the quoted fluxes and 
luminosities. } 
\end{deluxetable*}

We adapted the Princeton/MIT SDSS Spectroscopy routines\footnote{\url 
http://spectro.princeton.edu/idlspec2d\_doc.html}
to calculate redshifts. This technique cross-correlates
the spectra in pixel space with template spectra, with each pixel weighted by
the inverse of its variance, and is similar to the technique used in 
\citet{martini06}. The template spectra include a set of four eigenspectra for
galaxies, four eigenspectra for quasars, and forty eigenspectra for stars. The
five best galaxy redshifts for $-0.01<z<1.00$, five best quasar redshifts for
$0.0033<z<7.00$, and forty different stellar redshifts for $-0.004<z<0.004$ are
found and ordered by the reduced $\chi^2$ of their fit. We adopted the best-fit
redshift and classification for each source. To ascertain the quality of the fit
and the errors to the redshift, we resampled each spectra 100 times randomly
according to its noise characteristics and reran the cross-correlation routine.
Both the dispersion in best-fit redshifts and the best-fit spectral type were
used to qualify the spectral classification quality. 
If the dispersion in redshift was relatively low ($\sigma_z \lesssim 0.01$), 
$>68\%$ of the best-fit redshifts were within $3\sigma_z$ of our adopted 
redshift, and had the same spectral type (i.e., galaxy, quasar, or a similar 
stellar type) we consider this a secure redshift. 
Typically the maximum SNR of these spectra were $>5 {\rm \, pixel^{-1}}$.

We did not identify any AGN in these clusters with $L_{X,H} \geq 10^{43}$ 
\ergs, although we did identify several lower-luminosity AGN in these 
clusters. Data for the lower-luminosity X-ray sources are provided in 
Table~\ref{tbl:xcagn} and include several sources with spectroscopic 
measurements from the literature. The spectroscopic observations of Abell~1240 
and MS1512.4+3647 are complete for all candidates that would have 
$L_{X,H} \geq 10^{42}$ if at the cluster redshift, while the other four 
clusters are not complete to this luminosity limit. We have also measured 
redshifts, $R-$band magnitudes, and X-ray fluxes and luminosities for numerous 
additional sources not associated with these clusters and their properties are 
listed in Table~\ref{tbl:xsources}. 
As for the high-redshift clusters, several of the low-redshift clusters do 
not have direct velocity dispersion measurements. For Abell~1942 we estimated 
this quantity from the X-ray temperature. For Abell~1240 \citet{xue00} quote 
$kT = 3.83$ keV from \citep{mushotzky97}, but in fact the value in 
\citet{mushotzky97} appears instead to be for Abell 1242. As we could not 
identify another $T_X$ value in the literature, we used the measurement of 
$L_{bol} = 2.71 \times 10^{44}$ \ergs\ from \citet{david99} and the relation 
$\sigma = 10^{2.76} L_X^{0.19}$ derived by \citet{xue00} to estimate 
the velocity dispersion. 

\begin{deluxetable*}{llcccrrrrrr}
\tablecolumns{10}
\tabletypesize{\footnotesize}
\tablewidth{7.0truein}
\tablecaption{Nonmember X-ray Sources \label{tbl:xsources}}
\tablehead{
\colhead{CXOU ID} &
\colhead{$z$} &
\colhead{$z$ ref} &
\colhead{$R$} &
\colhead{$R$ flag} &
\colhead{$f_{X,S}$} &
\colhead{$f_{X,H8}$} &
\colhead{$f_{X,B}$} &
\colhead{log $L_{X,B}$} &
\colhead{log $L_{X,H}$} \\
\colhead{(1)} &
\colhead{(2)} &
\colhead{(3)} &
\colhead{(4)} &
\colhead{(5)} &
\colhead{(6)} &
\colhead{(7)} &
\colhead{(8)} &
\colhead{(9)} &
\colhead{(10)}
}
\startdata
J112314.9+431208.3 & $0.08017\pm0.00010$  & 1 & $17.66\pm0.08$ & 0 & $8.4^{+1.5}_{-1.3}$    & $29.9^{+4.7}_{-4.1}$ & $30.1^{+3.4}_{-3.1}$  & $41.69^{+0.05}_{-0.04}$ & $41.50^{+0.05}_{-0.04}$ \\
J112357.4+431314.1 & 0.08007              & 2 & $19.46\pm0.08$ & 0 & $23.8^{+2.4}_{-2.2}$   & $32.8^{+4.9}_{-4.3}$ & $55.9^{+4.5}_{-4.2}$  & $44.51^{+0.03}_{-0.03}$ & $44.32^{+0.03}_{-0.03}$ \\
J112403.0+431330.6 & 1.1049               & 2 & $18.39\pm0.08$ & 0 & $22.2^{+2.5}_{-2.2}$   & $17.3^{+4.2}_{-3.6}$ & $44.5^{+4.4}_{-4.0}$  & $43.16^{+0.04}_{-0.04}$ & $42.98^{+0.04}_{-0.04}$ \\
J112413.1+430639.3 & $2.3666\pm0.0015$    & 1 & $19.80\pm0.08$ & 0 & $7.53^{+1.3}_{-1.1}$   & $7.56^{+2.5}_{-2.0}$ & $16.1^{+2.4}_{-2.1}$  & $44.73^{+0.07}_{-0.06}$ & $44.54^{+0.07}_{-0.06}$ \\
J143804.9+033752.6 & $0.29192\pm0.00030$  & 1 & $18.50\pm0.06$ & 0 & $5.22^{+1.1}_{-0.95}$  & $<4.0$               & $7.8^{+1.8}_{-1.5}$   & $42.32^{+0.10}_{-0.09}$ & $42.13^{+0.10}_{-0.09}$ \\
J143832.2+033506.0 & $1.0083\pm0.0051$    & 1 & $19.98\pm0.06$ & 0 & $65.7^{+3.2}_{-3.0}$   & $79.8^{+6.2}_{-5.8}$ & $149.3^{+6.0}_{-5.8}$ & $44.85^{+0.02}_{-0.02}$ & $44.66^{+0.02}_{-0.02}$ \\
J143833.0+033606.8 & $0.38252\pm0.00017$  & 1 & $19.40\pm0.06$ & 0 & $11.5^{+1.3}_{-1.2}$   & $18.3^{+3.1}_{-2.7}$ & $28.5^{+2.6}_{-2.4}$  & $43.15^{+0.04}_{-0.04}$ & $42.96^{+0.04}_{-0.04}$ \\
J143839.7+033631.3 & $2.1493\pm0.0019$    & 1 & $19.00\pm0.06$ & 0 & $16.1^{+1.7}_{-1.5}$   & $16.1^{+3.1}_{-2.6}$ & $34.8^{+3.1}_{-2.8}$  & $44.97^{+0.04}_{-0.04}$ & $44.79^{+0.04}_{-0.04}$ \\
J143841.9+034110.2 & 1.7372               & 2 & $17.82\pm0.06$ & 0 & $28.2^{+2.3}_{-2.2}$   & $29.5^{+4.3}_{-3.8}$ & $61.6^{+4.3}_{-4.1}$  & $45.01^{+0.03}_{-0.03}$ & $44.83^{+0.03}_{-0.03}$ \\
J143847.3+032950.8 & $0.00034\pm0.00012$  & 1 & $16.89\pm0.06$ & 0 & $16.3^{+2.0}_{-1.8}$   & $2.0^{+2.2}_{-1.6}$  & $26.9^{+3.3}_{-3.0}$  & & \\
J143859.0+033547.8 & 0.7339               & 2 & $18.51\pm0.06$ & 0 & $46.5^{+2.8}_{-2.7}$   & $7.1^{+6.1}_{-5.7}$  & $113.8^{+5.6}_{-5.3}$ & $44.41^{+0.02}_{-0.02}$ & $44.22^{+0.02}_{-0.02}$ \\
J145623.0+221833.5 & $0.00027\pm0.00010$  & 1 & $15.51\pm0.07$ & 0 & $9.0^{+1.5}_{-1.3}$    & $5.1^{+2.5}_{-2.0}$  & $17.1^{+2.6}_{-2.3}$  & & \\
J145624.5+222057.1 & $0.00019\pm0.00010$  & 1 & $15.45\pm0.07$ & 0 & $14.8^{+1.4}_{-1.3}$   & $11.3^{+2.5}_{-2.2}$ & $29.8^{+2.5}_{-2.4}$  & & \\
J145634.6+221514.2 & $0.40918\pm0.00010$  & 1 & $20.16\pm0.07$ & 0 & $25.8^{+1.7}_{-1.6}$   & $56.8^{+4.4}_{-4.1}$ & $73.3^{+3.7}_{-3.5}$  & $43.63^{+0.02}_{-0.02}$ & $43.45^{+0.02}_{-0.02}$ \\
J145657.7+221315.6 & $0.00016\pm0.00010$  & 1 & $14.87\pm0.07$ & 0 & $8.63^{+1.0}_{-0.93}$  & $3.6^{+1.4}_{-1.1}$  & $16.0^{+1.8}_{-1.6}$  & & \\
J145708.7+222352.4 & 0.1238               & 2 & $17.44\pm0.07$ & 0 & $2.27^{+0.60}_{-0.50}$ & $<3.4$               & $3.32^{+1.0}_{-0.86}$ & $41.14^{+0.13}_{-0.11}$ & $40.95^{+0.13}_{-0.11}$ \\
J145710.7+221844.9 & $1.885\pm0.0014$     & 1 & $18.73\pm0.07$ & 0 & $3.99^{+0.66}_{-0.57}$ & $5.3^{+1.4}_{-1.1}$  & $9.4^{+1.2}_{-1.1}$   & $44.28^{+0.06}_{-0.05}$ & $44.09^{+0.06}_{-0.05}$ \\
J145712.3+221446.7 & $-0.00069\pm0.00010$ & 1 & $15.15\pm0.07$ & 1 & $50.4^{+2.1}_{-2.1}$   & $15.3^{+2.2}_{-2.0}$ & $90.2^{+3.6}_{-3.5}$  & & \\
J145721.0+222334.5 & $1.7362\pm0.0010$    & 1 & $19.33\pm0.07$ & 1 & $9.4^{+1.1}_{-1.0}$    & $7.0^{+1.9}_{-1.6}$  & $18.9^{+2.0}_{-1.9}$  & $44.50^{+0.05}_{-0.04}$ & $44.32^{+0.05}_{-0.04}$ \\
J145726.9+221755.1 & $1.4664\pm0.0011$    & 1 & $19.55\pm0.07$ & 0 & $23.6^{+1.7}_{-1.6}$   & $33.0^{+3.4}_{-3.1}$ & $56.3^{+3.2}_{-3.1}$  & $44.81^{+0.03}_{-0.02}$ & $44.62^{+0.03}_{-0.02}$ \\
J151427.0+363803.1 & 0.1616               & 2 & $16.90\pm0.06$ & 0 & $2.28^{+0.61}_{-0.49}$ & $1.8^{+1.9}_{-1.1}$  & $4.82^{+1.2}_{-0.99}$ & $41.53^{+0.11}_{-0.09}$ & $41.35^{+0.11}_{-0.09}$ \\
J151428.4+363743.5 & 0.4026               & 3 & $20.13\pm0.06$ & 0 & $7.70^{+1.0}_{-0.92}$  & $14.1^{+3.7}_{-3.0}$ & $18.9^{+2.2}_{-2.0}$  & $43.01^{+0.05}_{-0.05}$ & $42.83^{+0.05}_{-0.05}$ \\  
J151437.5+364041.3 & 0.1468               & 3 & $19.86\pm0.06$ & 0 & $12.1^{+1.2}_{-1.1}$   & $13.5^{+3.5}_{-2.9}$ & $26.9^{+2.4}_{-2.3}$  & $42.19^{+0.04}_{-0.04}$ & $42.01^{+0.04}_{-0.04}$ \\
J153938.1+662102.4 & 0.4375               & 4 & $19.71\pm0.08$ & 0 & $5.02^{+1.2}_{-0.98}$  & $6.5^{+2.9}_{-2.4}$  & $11.7^{+2.3}_{-2.1}$  & $42.90^{+0.09}_{-0.08}$ & $42.71^{+0.09}_{-0.08}$ \\
J154012.3+661439.2 & $1.0577\pm0.0029$    & 1 & $19.75\pm0.08$ & 0 & $37.2^{+1.9}_{-1.8}$   & $41.9^{+3.7}_{-3.4}$ & $83.1^{+3.7}_{-3.5}$  & $44.64^{+0.02}_{-0.02}$ & $44.46^{+0.02}_{-0.02}$
\enddata
\tablecomments{{\it Chandra} Observation Log. Columns are:
Col (1) Name of X-ray source;
Col (2) Redshift
Col (3) References for redshift are: 1: this work; 2: SDSS \citep{adelman-mccarthy08}; 3: \citet{abraham98}; 4: \citet{miller04};
Col (4) $R-$band magnitude;
Col (5) Flags for photometry are: (0) no flag; (1) may be contaminated by 
nearby neighbors or bad pixels;
Cols (6--8) Soft [0.5--2 keV], Hard [2--8 keV], and Broad [0.5--8 keV] band 
flux in the observed frame in units of 
$10^{-15} {\rm \, erg \, s^{-1} cm^{-2}}$. Upper limits are $3\sigma$ limits. 
Cols (9--10) Log of the Broad [0.5--8 keV] and Hard [2--10 keV] band luminosity 
in the rest-frame in units of \ergs\ corrected for Galactic absorption. We do 
not quote luminosities for X-ray sources identified with Galactic stars 
($z\sim0$). 
}
\end{deluxetable*}

\section{Cluster X-ray AGN Fraction} \label{sec:fa} 

We require two quantities to estimate the AGN fraction in these clusters:  
the number of AGN above our hard X-ray luminosity threshold hosted by galaxies 
with $M_R < M_R^*(z)+1$ and the total number of cluster galaxies above this 
magnitude threshold. For our low-redshift 
cluster sample, we have complete data to our X-ray threshold and 
reasonably complete data for the other cluster galaxies for about half of the 
clusters. For the high-redshift sample we have incomplete knowledge of both 
quantities. The AGN sample is likely incomplete because of spectroscopic 
incompleteness in the ChaMP and SEXSI surveys. The census of other cluster 
galaxies is very incomplete because few very high redshift clusters have the 
same quality membership data as our low-redshift sample. In the first three 
subsections below we describe the choice of the fiducial absolute magnitude 
threshold, our estimate of the completeness of the spectroscopic observations 
of X-ray sources, and the total number of cluster galaxies in the clusters 
with incomplete membership data. The fourth subsection describes our main 
result, the measurement of the AGN fraction and its evolution. The final 
two subsections describe potential contamination by AGN associated with 
large-scale structure around these clusters and other sources of 
uncertainty, respectively. 

\subsection{Host galaxy magnitude threshold} \label{galaxy} 

In previous work we defined the AGN fraction in clusters relative to 
galaxies more luminous than an $R-$band absolute magnitude of $M_R = 
-20$ mag \citep[\eg][]{martini06}. This choice of magnitude threshold was 
largely driven by expedience, namely it corresponded to the completeness 
limit for the most distant clusters in that sample. To properly extend this 
work to high redshift it is important to account for the evolution of the 
galaxy population in clusters, both in luminosity and number. These were 
not significant effects in our low-redshift study as the highest-redshift
cluster was at only $z=0.31$, but in our previous work at $z\sim0.6$ by 
\citet{eastman07} the $M_R = -20$ mag cutoff corresponded to a fainter 
absolute magnitude relative to $M_R^*$. Because the cluster galaxy population 
is larger, this would have led to a lower estimate of the AGN fraction if the 
cluster AGN are predominantly associated with the most luminous galaxies, as 
is the case at low redshifts \citep{sivakoff08}. 

Here we adopt an absolute magnitude threshold of $M_R^*(z) + 1$, 
and thus allow for evolution of $M_R^*$. At low-redshifts ($0.01 < z < 0.07$)
\citet{christlein03} measured the $R-$band luminosity function (LF) for six 
nearby clusters\footnote{Two of these clusters (Abell~85 and Abell~754) are 
in our low-redshift sample \citep{sivakoff08}.} and found that the composite 
cluster LF is consistent with a Schechter 
function with $M_R^* = -21.92 \pm 0.17$ mag ($h=0.7$, $\alpha = -1.21$).  
They also find an essentially identical value of $M_R^* = -21.93$ 
mag for the field. The low-redshift value of $M_R^* + 1$ is therefore 
about one magnitude brighter than the value of $M_R = -20$ mag we adopted in 
our previous, low-redshift studies \citep{martini06,sivakoff08}. For 
comparison, \citet{blanton03} measured $M^* = -21.22$ ($\alpha = -1.05$) 
at $z=0.1$ for the $r^{0.1}$ band on the AB system. This corresponds to 
$M_R^* = -21.72$ mag on the Vega system for the $R-$band at $z=0$ based on 
the conversions presented in \citet{blanton07} and is therefore consistent with 
\citet{christlein03}. 

Many recent studies have measured the evolution of $M_R^*$ and generally 
these measurements include both a value for all galaxies and separate 
measurements for particular spectroscopic types. This has relevance for 
our study as the cluster galaxy population is on average more quiescent than 
field galaxies and consequently their evolutionary history is different. 
We are most interested in measurements of the evolution of $M_R^*$ as a 
function of spectral type to isolate the evolution of galaxies dominated 
by older stellar 
populations that are most likely representative of the evolution of cluster 
galaxies.  A useful, low-redshift benchmark for a type-dependent LF for 
clusters comes again from \citet{christlein03}. They find $M_R^* = -21.78$ mag 
for quiescent galaxies in clusters, which is nearly identical to the 
value for all cluster members. For field galaxies \citet{chen03} use 
photometric redshifts in the Las Campanas Infrared Survey 
and measure values of $-21.70$ to $-22.22$ mag ($\alpha = -1$) for all galaxies 
over the range $0.5 < z < 1.5$ and values of $-21.21$ to $-21.82$ mag 
($\alpha = -0.2$) for galaxies consistent with an E/S0 + Sab spectral 
template. \citet{wolf03} use photometric redshifts from COMBO-17 and measure 
more pronounced evolution for their early-type spectral template with 
$M_R^*$ fading by $\sim 1$ mag from $z \sim 1.1$ to $z \sim 0.3$. 
More recently, \citet{ilbert05} measure a fading of $1.1 - 1.8$ mag between 
$z\sim 2$ and $z \sim 0.1$ in the $R-$band based on spectroscopic redshifts, 
although they do not present the evolution as a function of spectral type. 
These measurements of evolution in $M_R^*$ are broadly comparable to the 
1.2 mag of fading from $z=1$ to the present expected from pure luminosity 
evolution of a single stellar population with $z_f = 2$ and solar metallicity 
\citep{bruzual03}.  

Direct measurements of evolution of the cluster LF have mostly been conducted 
in the rest-frame $B-$band. \citet{goto05} find $M_B^* = -21.13$ mag for 
MS1054-03 ($z=0.83$), which is in our sample, and similar to the 
$M_B^* = -21.15$ mag measured for three clusters at an average $z = 0.859$ 
by \citet{postman01}. In comparison to local $B-$band measurements of the 
cluster LF \citep[e.g.][]{colless89,rauzy98}, \citet{goto05} conclude that 
$M_B^*$ fades by $0.46$ to $0.71$ mag between $z=0.83$ and $z=0$. For the 
same simple stellar population model considered above \citep{bruzual03}, 
1.2 mag of fading in $B-$band is expected from $z=0.83$ to the present. 
While there is not a direct measurement in the rest-frame $R-$band for the 
cluster LF, at yet longer wavelengths \citet{ellis04} find that the fading 
in the $K-$band is 1.2 mag from $z=0.9$ to the present and consistent with 
passive evolution and a formation epoch at $z_f=2$. From these investigations 
of the LF evolution in the field and clusters, we adopt the assumption that 
$M_R^*(z) = M_R^*(0) - z$ and the normalization for $M_R^*$ from 
\citet{christlein03} for all cluster galaxies to estimate the completeness of 
the spectroscopy 
of X-ray counterparts and the size of the galaxy population in low-redshift 
clusters. This result is broadly consistent with all of the results described
here, although is most consistent with the studies that predict more 
fading. If there is less fading of galaxies at the bright end of the LF, such 
as may be due to some low-level star formation in these galaxies, then 
the completeness limits we describe next are too bright and we will have 
systematically underestimated the population of luminous AGN in the 
higher-redshift clusters. 

\subsection{Completeness} \label{sec:completeness} 

We calculate a completeness limit in the observed $R-$band for each cluster 
based on the value of $M_R^*(z) + 1$ and a $k-$correction derived from 
the elliptical template of the four-component spectral template presented by 
\citet{assef08}. 
These templates are derived from 16,033 galaxies with spectroscopic redshifts 
and multiband photometry from the AGN and Galaxy Evolution Survey. Most
of the galaxies are in the range $0 < z < 1$ and the median redshift is 0.31. 
The parent sample is therefore broadly representative of our redshift range. 
For the higher-redshift clusters the $k-$correction requires a substantial 
extrapolation from the observed $R-$band, which for example samples rest-frame 
$B-$band at $z=0.5$. Our assumption that the typical cluster galaxies are
best approximated by an elliptical template is certainly reasonable for the 
low-redshift clusters. This may not be as good an approximation at higher
redshifts, although in a study of the color-magnitude relation in our two 
highest-redshift clusters (Lynx E and W) \citet{mei09} found there is no 
evidence for significant evolution. If a later-type template were a better
choice for the $k-$correction at higher redshift, the $k-$correction 
would be smaller and the necessary $R-$band spectroscopic limit would be 
brighter. The net effect would be a smaller completeness correction. 

\begin{figure}
\plotone{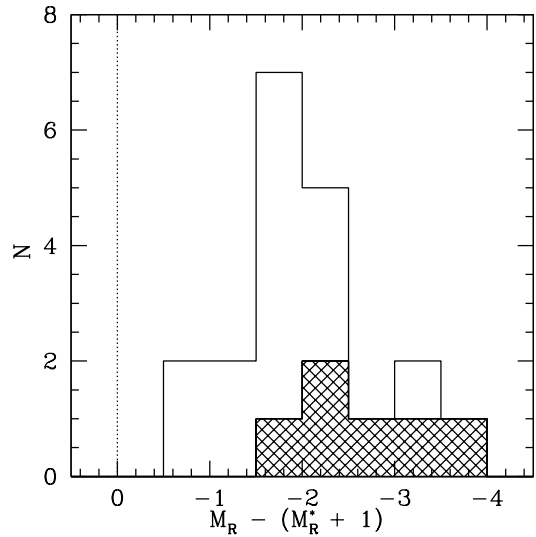}
\caption{Distribution in absolute magnitude $M_R$ of the cluster AGN 
relative to $M_R^*(z)+1$ at their redshift. All of the cluster AGN are 
substantially brighter than $M_R^*(z)+1$, although in most cases the 
spectroscopy is complete to this limit. The subset that are classified as 
BLAGN are represented by the hatched histogram. The dotted line 
corresponds to our galaxy luminosity threshold at $M_R^*(z)+1$. 
\label{fig:mr}
}
\end{figure}

The spectroscopic completeness of the high-z AGN sample largely depends on 
the completeness of the ChaMP and SEXSI surveys, although we also use 
additional spectra for MS 2053.7-0449, MS 1054-03, and RDCS J0910+5422. 
The ChaMP survey quotes a spectroscopic completeness of 77\% for $R<22.37$ mag 
\citep{silverman05a} and the SEXSI survey quotes a spectroscopic 
completeness of 61\% for sources with $22 < R < 23$ mag, 67\% for 
sources with $23 < R < 24$, and 74\% for sources with $R>24$ mag (typically 
to 24.4 mag) \citep{eckart06}. For the ChaMP data we adopt 77\% as the 
completeness correction for $R<22.37$ mag, while for the SEXSI survey we adopt 
an average completeness correction of 67\% for $R<24.4$ mag. For nearly all 
of the clusters above $z>0.6$ the spectroscopic data do not extend to the 
equivalent of $M_R^*(z) + 1$ and the size of the magnitude range without 
spectra ranges from a few tenths to over a magnitude. To estimate the 
number that may have been missed we inspected the host galaxy absolute 
magnitude distribution of the $L_{X,H} \geq 10^{43}$ \ergs\ AGN in the 
clusters with complete data and find only one AGN fainter than $M_R^*$. 
The distribution in $M_R$ of the X-ray AGN is shown in Figure~\ref{fig:mr}. 
We therefore assume that we have not missed any AGN because the spectroscopic 
observations of X-ray sources did not have the requisite depth, although this 
assumption may have led us to underestimate the AGN fraction at high 
redshift. In contrast, if our assumption of an early-type template for the 
$k-$correction was too red, then the spectroscopic data do achieve the 
requisite depth and this remains a nonissue. At brighter apparent magnitudes 
we do apply a completeness correction to account for the quoted 77\% and 
67\% completeness of the surveys. We discuss this further in 
\S\ref{sec:evolve} below. 


The X-ray AGN populations of several of these clusters have been studied  
in previous work. The first substantial study of spectroscopically-confirmed 
X-ray AGN in a high-redshift cluster was by \citet{johnson03} in MS1054-03. They 
identified 2 AGN associated with this cluster: CXOU J105702.7-033943 and 
CXOU J105710.6-033500; however, neither of these are included in the present 
sample. The first was not included because the X-ray luminosity is below our 
threshold of $10^{43}$ \ergs\ and the second because it falls slightly outside 
the projected virial radius ($R/R_{200} = 1.2$). \citet{martel07} have 
also studied X-ray sources in clusters, including three clusters that overlap 
this sample. They are discussed further in \S\ref{sec:hosts} below. 

\subsection{Inactive Cluster Galaxy Population} \label{sec:richness}

To estimate the AGN fraction in these clusters we need to know the 
number of cluster galaxies more luminous than $M_R^*(z) + 1$. We 
estimate this quantity in two ways, depending on the available data 
for the clusters. For the low-redshift clusters in our previous studies
\citep{martini06,martini07,sivakoff08} we have a large number of 
spectroscopically-confirmed cluster members and can estimate the number 
of cluster galaxies either directly or with a completeness correction. 
We have calculated new estimates for these clusters for the present paper
because we no longer use the $M_R=-20$ mag threshold of the previous 
work. These values are listed in Table~\ref{tbl:fa}. 

For essentially all of the new clusters in the present study we employ 
the same technique as \citet{eastman07} to estimate the number of 
cluster members above $M_R^*(z) + 1$ from the cluster velocity dispersion. 
This employs the richness--velocity dispersion relationship 
defined by \citet{koester07} for the MaxBCG cluster sample. The cluster 
richness $N_{gal}^{R200}$ is the number of red (E/S0) cluster members more 
luminous than $0.4L^*$ within the projected $R_{200}$ radius. This relationship 
was originally derived from a sample of 13,823 clusters with $0.1 < z < 0.3$ 
in the SDSS with velocity dispersions greater than $\sim 400$ \kms. 
\citet{becker07} provide the most recent estimate of this relation based on 
a larger sample that extends over both a broader redshift range and to lower 
velocity dispersion groups. They find $\ln \sigma = (6.17 \pm 0.04) + 
(0.436 \pm 0.015) \ln N_{gal}^{R200}/25$. For reference a 520 \kms\ cluster 
has $N_{gal}^{R200} = 30$. 

There are several caveats that need to be considered with the use of this 
estimator. 
First, the richness--velocity dispersion relationship is based on photometric 
and not spectroscopic redshifts. This is not a significant concern because 
for red cluster galaxies the photometric redshift estimates are robust within 
the quoted uncertainties. The second concern is that this relationship is 
based on the red cluster galaxies alone. At low redshifts this estimate is a 
reasonable approximation as the vast majority of cluster galaxies more 
luminous than $M_R^* + 1$ fall in this category. For example, the fraction 
of quiescent galaxies above this luminosity in the composite LF of 
\citet{christlein03} is $\sim 85$\%. While their definition of quiescence is 
based on spectral lines rather than color, these two definitions of 
quiescence typically agree when averaged over a cluster. At higher redshifts 
a larger fraction 
of the cluster galaxies may be blue due to ongoing star formation, but this 
can not be a substantial contribution because the luminosity-weighted mean 
star formation epoch is $z = 2$ for early-type cluster galaxies up to $z=0.5$ 
\citep{vandokkum07}. \citet{becker07} do find evidence of evolution 
in this relationship in the sense of lower richness at fixed velocity 
dispersion in higher redshift clusters, but they note that this may be due 
to their strict color selection. In addition, for our accounting of the 
inactive galaxy population the color of the galaxies does not matter so 
long as they are in the cluster and above the luminosity threshold. 
Observations of individual 
clusters with extensive spectroscopic data support the assumption that 
there is no substantial evolution in the relation between halo occupation 
number and cluster mass \citep{muzzin07b}. This is also supported by 
several theoretical studies that find minimal evolution in the number of 
bright galaxies in massive halos \citep{kravtsov04,zentner05}. 


\begin{figure}
\plotone{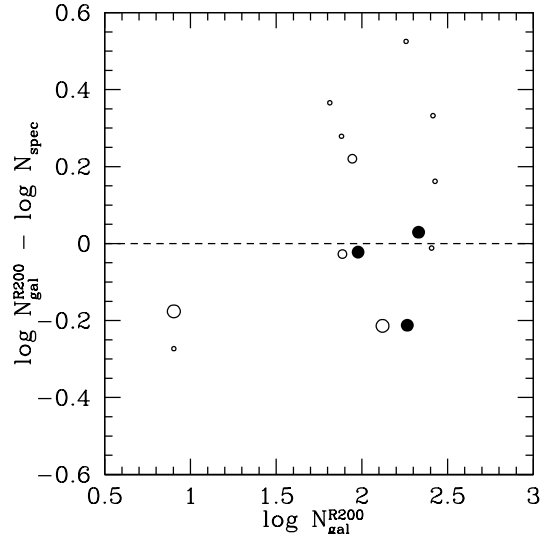} 
\caption{Difference between predicted and measured cluster richness compared 
to the cluster richness predicted by the MaxBCG sample. The quantity 
$N_{gal}^{R200}$ is the number of red cluster galaxies more luminous than 
$0.4L^*$ and estimated from the cluster velocity dispersion \citep{becker07}, 
while $N_{spec}$ is a spectroscopic estimate of this quantity (see 
\S\ref{sec:richness}). Symbols are coded according to the spectroscopic 
completeness relative to $R_{200}$. Large circles have complete coverage 
to $R_{200}$, medium circles have more than 50\% coverage, and the 
small circles
have less than 50\% coverage. Most clusters are at $z<0.5$ ({\it open 
symbols}), although substantial data exist for three at $z>0.5$ ({\it filled 
symbols}). See \S\ref{sec:richness} for further details. 
\label{fig:rich}
}
\end{figure}

We performed an independent validation of the MaxBCG relation with an analysis 
of the individual clusters in our sample with substantial membership data. 
While most 
of the low-redshift clusters have substantial membership data, these data 
generally do not extend to our estimate of $R_{200}$ \citep{martini07}, 
nor is the X-ray coverage complete to this radius. Our spectroscopic coverage 
was often limited to the size of the \chandra\ field of view. However, two 
useful exceptions are Abell 89B and MS1008.1-1224 and in both cases 
estimates agree to within a factor of two. Our wide-field X-ray coverage of 
Abell 85 and Abell 754 \citep{sivakoff08} were designed to sample a 
substantial fraction of the projected $R_{200}$ and these values also agree 
well. Figure~\ref{fig:rich} illustrates the difference between the MaxBCG 
membership estimates and our spectroscopic estimates. The larger points 
have nearly complete spectroscopic coverage to $R_{200}$, while smaller 
points are substantially more incomplete. These points indicate that the 
error introduced by adopting the MaxBCG relation is approximately a 
factor of two. This error estimate is also consistent with an examination 
of figure 4 of \citet{becker07}. 

At higher redshifts three of our clusters have extensive membership 
information. We estimate that MS0015.9+1609 has $\sim 200$ members based on 
several studies \citep{dressler92,ellingson98} and that MS2053.7-0449
has $\sim 100$ members \citep{tran05}. Note that these 
estimates are different from those presented in \citet{eastman07} due to 
updated completeness corrections and the change in the absolute magnitude 
threshold. For MS 1054-03 we estimate that there are $\sim 300$ members from 
the extensive spectroscopic work of \citet{tran07}. These three clusters 
are also shown in Figure~\ref{fig:rich} ({\it filled circles}). They 
are consistent with the low-redshift results and a factor of two uncertainty
in the richness -- velocity dispersion relation. While our estimates of 
the cluster galaxy population for these three clusters, as for the low-redshift
clusters, are based on all galaxies rather than just red galaxies, the 
consistency supports the assumption that the integral of the bright end of 
the galaxy luminosity function in clusters above an evolving $M_R$ threshold 
scales reasonably well with the cluster velocity dispersion independent of 
redshift, even if there is evolution in the colors of the cluster galaxies. 
The number of AGN, estimate of the inactive population, AGN fraction, 
and spectroscopic completeness for each cluster is listed in 
Table~\ref{tbl:fa}. 

\subsection{Cluster AGN Fraction and Evolution} \label{sec:evolve}


\begin{figure}
\plotone{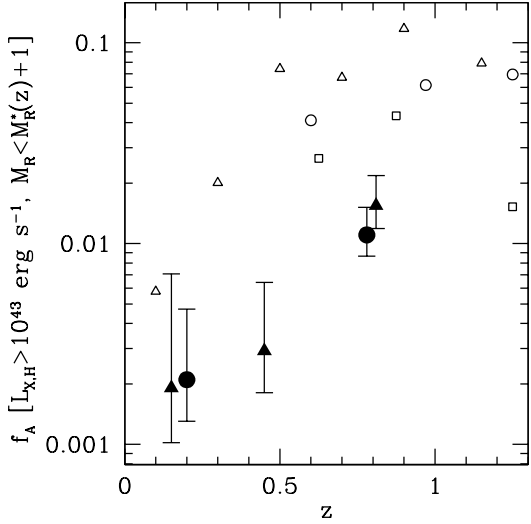} 
\caption{Evolution of the AGN population in clusters from $z=0$ to $z=1.3$
({\it filled symbols}). The fraction of cluster members more luminous than 
$M_R^* + 1$ with AGN that have $L_{X,H} > 10^{43}$ \ergs is shown in two 
redshift bins ($z<0.4$, $z>0.4$; {\it filled circles}) and three redshift bins 
($z<0.3$, $0.3<z<0.8, 0.8<z<1.3$; {\it filled triangles}). 
We also show our estimate of the field AGN fraction based on the galaxy LF 
estimates by \citet[][{\it open triangles}]{ilbert05}, \citet[][{\it open 
circles}]{dahlen05}, and \citet[][{\it open squares}]{chen03}. See 
\S\ref{sec:evolve} for further details. 
\label{fig:faz}
}
\end{figure}


\begin{deluxetable}{llrccclc}
\tablecolumns{8}
\tablewidth{0pt}
\tabletypesize{\scriptsize}
\tablecaption{AGN Fraction Estimates and Cluster Membership\label{tbl:fa}}
\tablehead{
\colhead{Cluster} &
\colhead{$z$} &
\colhead{$\sigma$} &
\colhead{$N_{AGN}$} &
\colhead{$N_{gal}$} &
\colhead{Flag} &
\colhead{$f_{A,raw}$ [\%]} &
\colhead{$f_{spec}$}  \\
\colhead{(1)} &
\colhead{(2)} &
\colhead{(3)} &
\colhead{(4)} &
\colhead{(5)} &
\colhead{(6)} &
\colhead{(7)} &
\colhead{(8)} 
}
\startdata
Abell754&	0.0546 &  953 & 1 &  82 & 1 & $1.2^{+2.8}_{-1.0}$ &    1.00 \\
Abell85&	0.0554 &  993 & 0 &  53 & 1 & $<2.2$  &                1.00 \\
Abell3128&	0.0595 &  906 & 0 &  28 & 1 & $<4.1$  &                1.00 \\
Abell3125&	0.0616 &  475 & 0 &  15 & 1 & $<7.7$  &                1.00 \\
Abell644&	0.0701 &  952 & 0 &  40 & 1 & $<2.9$  &                1.00 \\
Abell89B&	0.0770 &  474 & 0 &  12 & 1 & $<9.6$  &                1.00 \\
Abell2104&	0.1544 & 1242 & 1 &  54 & 1 & $1.9^{+4.3}_{-1.5}$ &    1.00 \\
Abell1240&	0.1590 &  698 & 0 &  28 & 2 & $<4.1$  &                1.00 \\
Abell1689&	0.1867 & 1400 & 0 & 184 & 1 & $<0.62$ &                1.00 \\
Abell2163&	0.2007 & 1381 & 0 & 262 & 1 & $<0.44$ &                1.00 \\
Abell1942&	0.2240 &  905 & 0 &  65 & 2 & $<1.8$  &                1.00 \\
Abell2125&	0.2465 & 1113 & 0 & 127 & 2 & $<0.90$ &                1.00 \\
MS1455.0+2232&	0.2578 & 1032 & 0 &  99 & 2 & $<1.2$  &                1.00 \\
MS1008.1-1224&	0.3068 & 1127 & 0 & 216 & 1 & $<0.53$ &                1.00 \\
AC114&		0.3148 & 1388 & 0 & 121 & 1 & $<0.95$ &                1.00 \\
ZwCl1358.1+6245&0.328  & 1003 & 0 &  91 & 2 & $<1.3$  &                1.00 \\
MS1512.4+3647&	0.372  &  575 & 0 &  15 & 2 & $<7.7$  &                1.00 \\
MS1621.5+2640&	0.430  &  735 & 0 &  65 & 2 & $<1.8$  &                0.67 \\
3C295&		0.460  & 1642 & 2 & 412 & 2 & $0.49^{+0.64}_{-0.31}$ & 0.67 \\
MS0451.6-0305&	0.538  & 1371 & 0 & 273 & 2 & $<0.42$ &                0.77 \\
MS0015.9+1609&	0.541  & 1234 & 1 & 214 & 2 & $0.47^{+1.1}_{-0.39}$ &  0.77 \\
RXJ0848.7+4456&	0.574  &  895 & 1 & 102 & 2 & $0.98^{+2.3}_{-0.81}$ &  0.67 \\
MS2053.7-0449&	0.583  &  865 & 0 &  95 & 2 & $<1.2$ &                 1.00 \\
RXJ0542.8-4100&	0.634  & 1269 & 5 & 229 & 2 & $2.18^{+1.5}_{-0.94}$ &  0.77 \\
RXJ2302.8+0844&	0.722  &  658 & 0 &  50 & 2 & $<2.3$ &                 0.77 \\
MS1137.5+6625&	0.782  &  885 & 1 & 100 & 2 & $1.00^{+2.3}_{-0.83}$ &  0.77 \\
RX J1317.4+2911&0.805  & 1142 & 1 & 179 & 2 & $0.56^{+1.3}_{-0.46}$ &  0.67 \\
RXJ1716.4+6708&	0.813  & 1445 & 3 & 308 & 2 & $0.97^{+0.95}_{-0.53}$ & 0.92 \\
MS 1054-03&	0.823  & 1156 & 1 & 184 & 2 & $0.54^{+1.2}_{-0.45}$ &  0.77 \\
RDCS J0910+5422&1.110  &  675 & 1 &  53 & 2 & $1.9^{+4.3}_{-1.6}$ &    0.67 \\
Lynx E&		1.261  &  740 & 1 &  66 & 2 & $1.5^{+3.5}_{-1.3}$ &    0.67 \\
Lynx W&		1.270  &  650 & 1 &  49 & 2 & $2.0^{+4.7}_{-1.7}$ &    0.67
\enddata
\tablecomments{
AGN fraction estimates for individual clusters. Columns are: 
Col. (1): Cluster name; 
Col. (2): Redshift; 
Col. (3): Velocity dispersion (references for these values are in 
Table~\ref{tbl:highz}, Table~\ref{tbl:lowz}, \citet{sivakoff08} for 
Abell 754, Abell 85, Abell 89B, \citet{martini06} for Abell 3128, Abell 3125, 
Abell 644, Abell 2104, Abell 2163, and MS1008.1-1224, or adopted from 
\citet{czoske04} for Abell 1689 and \citet{girardi01} for AC 114);  
Col. (4): Number of AGN with $L_{X,H} \geq 10^{43}$ \ergs in galaxies 
more luminous than $M_R^*(z) + 1$; 
Col. (5): Estimate of the number of cluster galaxies more luminous than 
$M_R^*(z) + 1$ within either the \chandra\ FOV or $R_{200}$, whichever is 
smaller;
Col. (6): Flag for the origin of the estimate where 1: from 
our spectroscopy and completeness correction; 2: from 
the MaxBCG as described in \S\ref{sec:richness}; 
Col. (7): Estimate of the cluster AGN fraction in percent; 
Col. (8): Estimate of the spectroscopic completeness for X-ray sources. 
}
\end{deluxetable}


\begin{deluxetable*}{llrlrrrlll}
\tablecolumns{10}
\tablewidth{7.0truein}
\tabletypesize{\scriptsize}
\tablecaption{AGN Fraction for Subsamples of the Clusters\label{tbl:fabin}}
\tablehead{
\colhead{Sample} &
\colhead{$z$ range} &
\colhead{$N_{CL}$} &
\colhead{median $z$} &
\colhead{median $\sigma$} &
\colhead{$N_{A,raw}$} &
\colhead{$N_{gal}$} &
\colhead{$f_{A,raw}$ [\%]} & 
\colhead{$f_{spec}$} & 
\colhead{$f_{A,corr}$ [\%]} \\ 
\colhead{(1)} &
\colhead{(2)} &
\colhead{(3)} &
\colhead{(4)} &
\colhead{(5)} &
\colhead{(6)} &
\colhead{(7)} & 
\colhead{(8)} & 
\colhead{(9)} 
}
\startdata
Two Bins	&	  	&    &      &    &      &                                &      &                         \\ 
		& $z<0.4$ 	& 17 & 0.19 & 993 &  2 & 1492 & $0.134^{+0.18}_{-0.087}$ & 1.00 & $0.134^{+0.18}_{-0.087}$ \\ 
  		& $z>0.4$ 	& 15 & 0.72 & 895 & 18 & 2379 & $0.76^{+0.22}_{-0.18}$   & 0.76 & $1.00^{+0.29}_{-0.23}$ \\ 
	 	&	  	&    &      &    &      &                                &      &                         \\ 
Three Bins	&	  	&    &      &    &      &                                &      &                         \\ 
     		& $z<0.3$ 	& 13 & 0.15 & 953 &  2 & 1049 & $0.19^{+0.25}_{-0.12}$   & 1.00 & $0.19^{+0.25}_{-0.12}$ \\ 
  		& $0.3<z<0.6$ 	& 10 & 0.45 &1065 &  4 & 1604 & $0.25^{+0.20}_{-0.12}$   & 0.81 & $0.31^{+0.24}_{-0.15}$ \\ 
		& $z>0.6$	&  9 & 0.81 & 885 & 14 & 1218 & $1.15^{+0.39}_{-0.30}$   & 0.78 & $1.47^{+0.50}_{-0.39}$
\enddata
\tablecomments{
Cluster AGN fractions with the data split into two bins and three bins. 
The two bins are split at $z=0.4$, while the three bins split the data 
at $z=0.3$ and $z=0.6$. For each bin we list: 
Col. (2): redshift range; 
Col. (3): number of clusters; 
Col. (4): median redshift; 
Col. (5): median velocity dispersion of clusters; 
Col. (6): sum of the luminous AGN in the bin; 
Col. (7): raw AGN fraction with double-sided, $1-\sigma$ confidence limits; 
Col. (8): estimate of the mean spectroscopic completeness weighted by the 
number of galaxies per cluster; 
Col. (9): AGN fraction corrected for spectroscopic completeness. 
}
\end{deluxetable*}

The AGN fraction for any single cluster is very small and it is uncertain due 
to small number statistics. In addition, the AGN fraction may vary from 
cluster to cluster due to correlations with other cluster properties such as 
velocity dispersion \citep[][]{sivakoff08}. The AGN fraction may also depend on 
variations in the properties of the galaxy population within each cluster 
(e.g., mass, SFR, morphology). We therefore have binned the 
cluster sample in two ways to characterize variations with redshift. First, we 
simply split the sample at $z=0.4$. This choice is primarily motivated by 
the transition between where we rely on our own measurements and where we 
largely rely on other work. It also approximately divides the sample in two 
(17 clusters at $z<0.4$, 15 at $z>0.4$). 
This yields completeness-corrected AGN fractions of $f_A(z=0.19) = 
0.00134^{+0.0018}_{-0.00087}$ and $f_A(z=0.72) = 0.0100^{+0.0029}_{-0.0023}$, 
or approximately a factor of eight increase in the AGN fraction (see 
Table~\ref{tbl:fabin}) from a median redshift of 0.19 to a median redshift of 
0.72.  AGN fractions without the completeness correction are also listed in 
Table~\ref{tbl:fabin}. The uncertainties on these quantities are double-sided, 
$1-\sigma$ confidence limits \citep{gehrels86}. The increase in the AGN fraction is 
formally significant at the $3.8\sigma$ level. We also split the 
sample into three bins with $z<0.3$, $0.3 < z < 0.6$, and $z>0.6$ to 
better resolve the continued increase at high redshift that is apparent 
in the raw data for individual clusters. This binning yields AGN fractions 
of $f_A(z=0.15) = 0.0019^{+0.0025}_{-0.0012}$, $f_A(z=0.45) = 
0.0031^{+0.0024}_{-0.0015}$, and $f_A(z=0.81) = 0.0147^{+0.0050}_{-0.0039}$. 
The measured evolution between the lowest and highest bins is also a factor of 
eight and in good agreement with the other binning scheme. We note that the 
observed evolution is also well fit by a simple power law scaling as 
$f_A \propto (1+z)^{\alpha}$ where $\alpha = 5.3^{+1.8}_{-1.7}$, although the 
power-law index is strongly correlated with the $z=0$ value of the AGN 
fraction. 

The factor of eight evolution of the AGN fraction is smaller but consistent 
with the order of magnitude evolution observed by \citet{eastman07}. They 
measured $f_A(z=0.2) = 0.0007^{+0.0021}_{-0.0007}$ and $f_A(z=0.6) = 
0.020^{+0.012}_{-0.008}$ for $L_{X,H} > 10^{43}$ \ergs, although for a lower 
and fixed galaxy absolute magnitude of $M_R = -20$. At $z=0$ our galaxy 
absolute magnitude threshold is approximately a magnitude brighter than that 
used by \citet{eastman07} and the offset increases linearly with redshift. 
This difference in absolute magnitude threshold can readily account for the 
change in the low-redshift fraction because most of the AGN are associated 
with luminous cluster galaxies, that is increasing the galaxy luminosity 
threshold decreases the denominator and does not affect the numerator of the 
AGN fraction. In addition, we have since identified a second luminous AGN at 
low redshift \citep{sivakoff08}. At high redshift the change in galaxy 
luminosity threshold is also important, but in addition the cluster sample is 
more than three times larger than the \citet{eastman07} sample. The 
low-redshift cluster sample has increased by less than a factor of two. 

One way to characterize the evolution of the cluster AGN fraction relative 
to the field is to calculate the integral of the field space density
$\Phi(L_{X,H}>10^{43})$ as a function of redshift. Integration of the 
luminosity-dependent density evolution model in \citet{ueda03} yields a 
factor of five increase between $z=0.8$ and $z=0.2$, which is somewhat less 
but consistent with the observed evolution of cluster AGN. 
However, this is not a fair comparison because the evolution of field 
AGN with $\Phi(L_{X,H}>10^{43})$ is not normalized by the evolution of all 
field galaxies brighter than $M_R^* + 1$ and the cluster AGN fraction is. 

While there is not a direct measurement of the field AGN fraction similar to 
our calculation for clusters \citep[although see][]{lehmer07}, we can estimate 
this quantity by dividing the integral of the field hard X-ray LF from 
\citet{ueda03} by the integral of the galaxy LF. We have identified three 
surveys that report LF measurements for the $R-$band and approximately span 
the same redshift range of this work. The first of these is the VIMOS-VLT Deep 
Survey \citep{ilbert05}, which is based on $UBVRI$ photometry, $\sim 11,000$ 
spectra to $I_{AB} = 24$ mag and extends from $z=0.05$ to $z=2$ \citep[although
their lowest-redshift point is taken from SDSS;][]{blanton03}. We also show 
results from two measurements based on photometric redshift data: the Las 
Campanas Infrared Survey \citep[LCIRS;][]{chen03}, which is mostly based on 
$UBVRIH$ measurements and presents the LF for $z=0.5-1.5$, and the Great 
Observatories Origins Deep Survey \citep[GOODS;][]{dahlen05}, which is based on 
$U$ through $K$ observations and presents the galaxy LF to $z=2$. While these 
photometric redshift surveys may have more systematic uncertainties than the LF 
based on spectroscopic measurements, they have the virtue that they have 
measured the luminosity function in the rest-frame $R-$band, rather than relied 
on assumptions about galaxy spectral energy distributions (SEDs) to calculate 
$k-$corrections. We have calculated the field AGN fraction for each of these 
surveys and show the results in Figure~\ref{fig:faz} ({\it open symbols}). At 
low redshift the AGN fraction calculated with the \citet{ilbert05} LF is 
approximately a factor of five above the cluster fraction, which is consistent 
with the difference between the field and clusters seen by \citet{dressler99} 
for spectroscopically-identified AGN. At higher redshifts ($z>0.5$), the field 
estimates range between a factor of three and a factor of ten above the cluster 
fraction. These estimates of the field AGN fraction vary so substantially due 
to the dispersion in estimates of the galaxy luminosity function. In addition, 
this calculation presupposes that all of the X-ray AGN are in galaxies more 
luminous than $M_R^*(z) + 1$. While there is good evidence that most of these 
luminous X-ray AGN are in relatively luminous galaxies 
\citep[e.g.][]{silverman09a}, there is nevertheless a bias against 
spectroscopic identification of lower-luminosity X-ray AGN host galaxies. 
Finally, we note that the relative evolution of galaxies in clusters and the 
field further complicates this comparison. In future work we hope to compile 
sufficient data to calculate the AGN fraction in the field and clusters as a 
function of galaxy mass. At present the data are insufficient to conclude 
if the cluster AGN fraction or field AGN fraction evolves more rapidly. 

\subsection{Contamination by AGN Associated with Large-Scale Structure} \label{sec:lss} 

One concern raised about the physical origin of the Butcher-Oemler effect 
is the contribution of projection effects. \citet{diaferio01} studied this 
issue in detail with N-body simulations and semianalytic models 
to distinguish true cluster members from field interlopers that were 
at the cluster redshift and within the projected $R_{200}$, yet 
physically outside the cluster $R_{200}$. \citet{diaferio01} concluded 
that up to 50\% of the apparent Butcher-Oemler galaxies at the 
redshifts of high-redshift clusters may be interlopers. A similar 
effect may be relevant for the AGN population and such a large contamination 
would decrease the observed evolution, but not erase it. 

While there is no comparable study that directly investigates the 
projection of AGN onto high-redshift clusters, there is good evidence 
that AGN are associated with the large-scale environment of clusters.  
\citet{gilmour07} identified 11 X-ray AGN (to a lower luminosity limit 
of $\sim 10^{41}$ \ergs) in the A901/2 supercluster at $z \sim 0.17$ and 
only one was in the densest region of the supercluster. The remainder were 
mainly in regions of intermediate density. In the vicinity of 3C295 
($z=0.46$) \citet{delia08} find evidence for AGN associated with a filamentary 
structure. At yet higher redshifts this trend is also 
apparent. \citet{kocevski09a} find X-ray AGN associated with the CL1604 
supercluster at $z\sim0.9$, which contains 8 confirmed groups and clusters. 
These AGN mostly avoid the densest regions of the clusters and are located 
on the outskirts of the most massive clusters, that is they are associated with 
poorer clusters and groups. 

We examined our data to determine if there were a population of 
AGN outside the projected $R_{200}$ for these clusters similar to 
those seen in the two superclusters. This is only possible with the 
subset of the sample with substantial coverage beyond $R_{200}$. 
Eight of the clusters have \chandra\ coverage that extends to 
$2 R_{200}$. There are six AGN between $R_{200}$ and $2 R_{200}$ that 
meet our velocity cuts for cluster membership compared to 
eight AGN within $R_{200}$ for these same clusters. 
The larger number within the clusters suggests the opposite trend from 
the two supercluster studies described above, although these 
results are not truly in conflict because the supercluster studies 
encompassed a much larger area outside of dense clusters than this 
study. The different large-scale environments associated with these clusters 
and the superclusters suggest a more quantitative comparison would not 
be meaningful. These large-scale structure data also provide a crude means 
to estimate the likelihood of chance juxtapositions of AGN associated with 
large-scale structure onto the clusters. If interloper AGN have the same 
surface density within $R_{200}$ as between $R_{200}$ and $2 R_{200}$, then 
the six we identified in an area of $3 \pi R_{200}$ suggest we should expect at 
most 2 interlopers compared to the 8 AGN we see within $R_{200}$. 
This line of argument suggests that the interloper fraction is 25\%, which is 
small compared to the observed evolution signature. 


\begin{deluxetable*}{llccccccl}
\tablecolumns{9}
\tablewidth{7.0truein}
\tabletypesize{\scriptsize}
\tablecaption{High-Redshift AGN Associated with Large-Scale Structure around Clusters\label{tbl:lssagn}}
\tablehead{
\colhead{AGN} &
\colhead{Cluster} &
\colhead{$z$} &
\colhead{$R$ [mag]} &
\colhead{log $L_{X,H}$ [\ergs]} &
\colhead{$\delta v/\sigma$ [km/s]} &
\colhead{$\Delta R$ [arcmin]} &
\colhead{$R/R_{200}$} & 
\colhead{Class} \\ 
\colhead{(1)} &
\colhead{(2)} &
\colhead{(3)} &
\colhead{(4)} &
\colhead{(5)} &
\colhead{(6)} &
\colhead{(7)} &
\colhead{(8)} & 
\colhead{(9)} 
}
\startdata
CXOSEXSI J084846.0+445945&	RX J0848.7+4456 &0.567	&21.45	&43.1	&1.99	&3.51	&1.16	&ELG  \\
CXOMP J230300.9+084659&		RXJ2302.8+0844	&0.738	&21.71	&44.23	&2.81	&4.46	&1.2	&BLAGN  \\
CXOSEXSI J171807.6+670647&	RXJ1716.4+6708	&0.797	&21.75	&44	&1.83	&7.8	&1.59	&BLAGN  \\
CXOU J105710.6-033500&		MS 1054-03	&0.832	&21.93	&43.14	&1.27	&4.57	&1.18	&ALG  \\
CXOSEXSI J091040.8+542006&	RDCS J0910+5422	&1.097	&22.38	&43.1	&2.74	&2	&1.13	&ELG  \\
CXOSEXSI J084903.9+445023&	LynxE		&1.276	&23.92	&43.2	&2.95	&1.76	&1.03	&ELG
\enddata
\tablecomments{
AGN associated with large-scale structure around the subset of high-redshift 
clusters with complete X-ray coverage to twice the projected virial radius. 
This is the subset of AGN that satisfy the redshift selection criterion, 
but have a projected distance of $1 < R/R_{200} \leq 2$. 
Columns are identical to Table~\ref{tbl:highzagn}. The data for 
CXOU J105710.6-033500 are from \citet{vandokkum00} for the redshift, 
magnitude, and classification and the X-ray data are from \citet{johnson03}. 
This sample is described in further detail in \S\ref{sec:lss}. 
}
\end{deluxetable*}

\subsection{Uncertainties} \label{systematics} 

One major potential source of systematic error is the use of the MaxBCG 
richness estimator to estimate the fraction of cluster galaxies more luminous 
than $M_R^*+1$. In \S\ref{sec:richness} we estimated that there is a factor of 
two uncertainty in the use of this relation. This uncertainty is mainly 
important for the high-redshift subsamples as the low-redshift subsamples 
have more complete spectroscopic membership data. If we randomly introduce a 
factor of two uncertainty in each cluster, the effect is negligible when 
averaged over the 15 clusters with $z>0.4$ compared to the factor of eight
evolution in the AGN fraction. 

As mentioned previously, another valid concern with the MaxBCG estimator is 
that it is calibrated to the number of red galaxies in the cluster and this 
population may not all be in place at $z=0.4$ and higher. For our application 
it does not matter if the galaxies are red or not, just that they are in the 
cluster. Furthermore, if we have overestimated the number of galaxies brighter 
than $M_R^*+1$ then we have underestimated the evolution of the AGN fraction 
and our result is yet more statistically significant. The assumption that 
all of the galaxies are red does impact the $k-$correction we use to estimate 
the spectroscopic limit corresponding to $M_R^*(z)+1$ and thus the size of 
our completeness correction. If the galaxies are redder, then the 
$k-$correction would be smaller, the apparent magnitude limit would be 
brighter, and the completeness correction would be smaller. The implication 
would be that we have preferentially overestimated the AGN fraction at high 
redshifts because completeness corrections are only applied to the 
high-redshift clusters. While the average completeness correction approaches 
25\% (see Table~\ref{tbl:fabin}), in practice the spectroscopic completeness 
is not a strong function of apparent magnitude \citep[e.g. see 
\S\ref{sec:completeness},][]{silverman05a,eckart06} and 
we consequently expect much less than a 25\% reduction in the evolution. 
The evolution of the host galaxy population is also important because 
if there were less fading of $M^*(z)$ than we assume, then the completeness 
limit would be too bright and we would have underestimated the AGN fraction 
at high redshift. 

The value of the cluster velocity dispersion introduces additional uncertainty 
to this calculation in two ways. First, many of the direct measurements of the 
cluster velocity dispersion, particularly for high-redshift clusters, are 
based on small samples of galaxies and thus the velocity dispersion itself 
may be uncertain, particularly if the galaxy velocity distribution is 
not Gaussian. Second, as noted above the cluster velocity dispersion has not 
been directly measured for several clusters and we instead used the X-ray 
temperature and the results of \citet{xue00} to estimate the velocity 
dispersion and this has a 30\% scatter. We checked both of these concerns 
with a measurement of the scatter between $\sigma$ and $T_X$ for 
the ten high-redshift clusters with measurements of both quantities 
and the mean deviation is $\sim 220$ \kms\ if we exclude 3C295, which 
has a substantially higher velocity dispersion \citep[1642 km/s][]{girardi01} 
than expected from its X-ray temperature \citep[5.3 K from][]{vikhlinin02}. 
This mean deviation corresponds to approximately a factor of two uncertainty 
in the richness, which is comparable to the uncertainty we derived for the 
richness estimator. From this analysis we similarly conclude that this 
source of uncertainty does not substantially affect our results. 

A related evolutionary effect is that the velocity distributions of the 
high-redshift clusters may be systematically more non-Gaussian than 
low-redshift clusters because the high-redshift clusters are less likely to be 
relaxed. If the cluster velocity dispersion were overestimated, then the 
richness and $R_{200}$ would be overestimated as well. This in turn would lead 
to an underestimate of the AGN fraction in high-redshift clusters. 
\citet{jeltema05} measured power ratios from \chandra\ observations of the IGM 
for a large sample of clusters out to $z\sim1$ and found good evidence that 
high-redshift clusters are less relaxed than low-redshift clusters, so this 
potential source of systematic error would lead us to underestimate the AGN 
fraction. Nine of our clusters were analyzed in the \citet{jeltema05} study, 
including eight in our high-redshift sample. We compared the AGN fractions and 
the power ratios for these clusters, but did not find a significant trend.  
Unfortunately we do not have sufficient redshift data for most high-redshift 
clusters to look for non-Gaussianity in the galaxy velocity distribution, 
although note there is no evidence for a trend between dynamically-disturbed 
clusters and AGN fraction at low redshift \citep{martini07}. 

\begin{figure}
\plotone{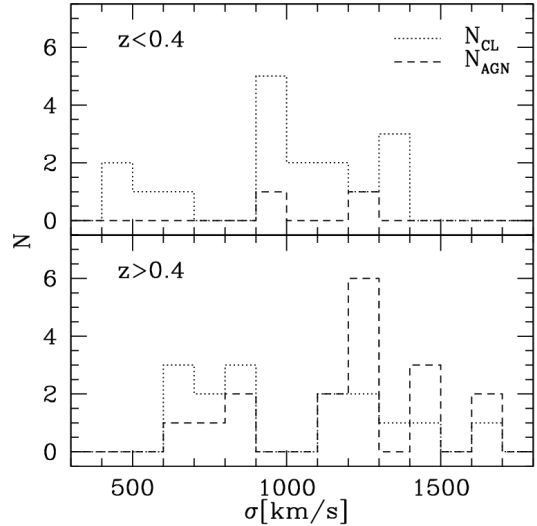} 
\caption{
Histograms of the number of clusters with a given velocity dispersion 
({\it dotted line}) and the number of AGN in clusters of a given velocity 
dispersion ({\it dashed line}) for the low-redshift ($z<0.4$; {\it top panel}) 
and the high-redshift ($z>0.4$; {\it bottom panel}) subsamples. The cluster
samples are reasonably well matched within these two redshift bins. 
\label{fig:sigma}
}
\end{figure}

Finally, we consider the evolution of the cluster population to determine if 
the higher-redshift clusters represent the progenitor population of the 
lower redshift clusters. As noted previously, observations at low redshift 
indicate that the AGN fraction depends on environment and specifically that 
the AGN fraction is higher in lower velocity dispersion environments 
\citep{sivakoff08,arnold09}. Therefore if our high-redshift clusters are the 
progenitors of lower velocity dispersion clusters or massive groups, then 
the observed evolution may not be as significant. As many of the high-redshift 
clusters are X-ray selected, they are generally high-mass clusters and are 
reasonably well matched to the lower-redshift sample 
(see Figure~\ref{fig:sigma} and Table~\ref{tbl:fa}).  
Following \citet{finn05} and \citet{poggianti06}, we have estimated the 
velocity dispersions of the progenitors of the high-redshift cluster sample 
and find they are in good agreement. For example, the progenitor of a 1000 
\kms\ cluster at $z=0$ has 800 \kms\ at $z=0.6$ \citep{poggianti06}, or only 
about 100 \kms\ less than the difference between our low-redshift and 
high-redshift subsamples. The sense of this trend is that the high-redshift 
sample is actually somewhat more massive than the typical progenitor of the 
low-redshift sample and therefore the minor mismatch in cluster masses 
is more likely to have dampened rather than enhanced the measured evolution 
of the AGN fraction. 

\section{Properties of the Cluster AGN} \label{sec:agn}

\subsection{Distribution} \label{sec:dist} 


\begin{figure*}
\plotone{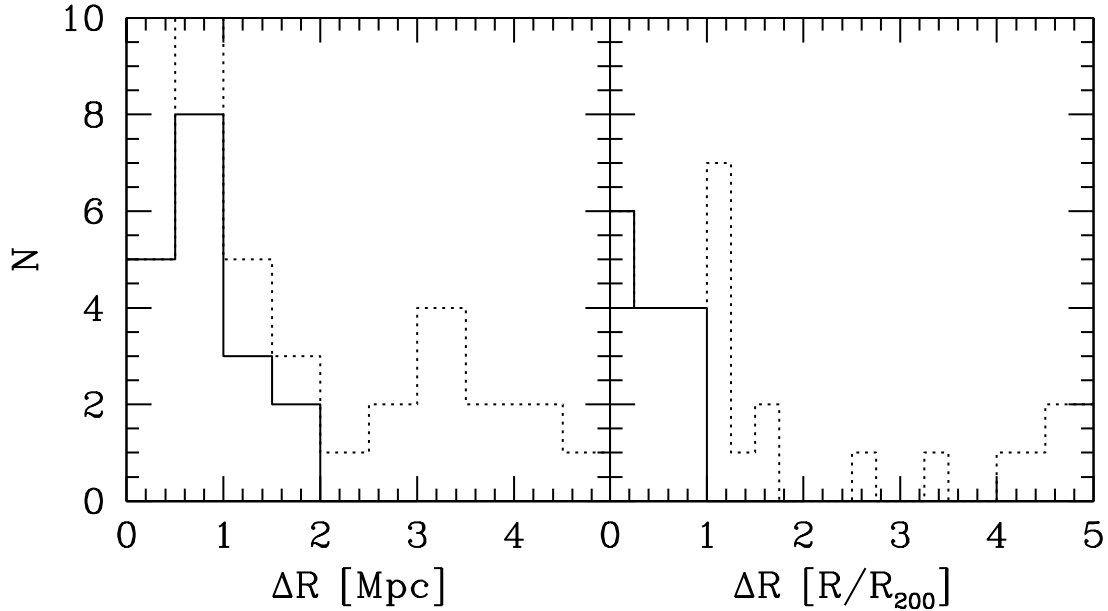} 
\caption{
Histograms of the AGN clustercentric distances in terms of Mpc ({\it left}) 
and normalized to $R_{200}$ ({\it right}) for cluster AGN with $z>0.4$. The 
distribution of the confirmed cluster members ({\it solid line}) is much more 
centrally peaked when expressed in terms of Mpc than in terms of $r/R_{200}$. 
Other AGN associated with large-scale structure (with $R>R_{200}$) are also 
shown ({\it dotted line}). 
\label{fig:rad}
}
\end{figure*}

The projected radial and velocity distributions of the AGN provide valuable 
additional information about the origin of the AGN. For example, if the 
AGN are preferentially located in the cluster outskirts, or preferentially 
have a higher velocity dispersion than the cluster mean, this may indicate 
that their host galaxies have relatively recently entered the cluster 
potential. This is known to be the case for emission-line galaxies 
\citep{biviano97,dressler99}. At low-redshifts and for lower-luminosity 
X-ray AGN, \citet{martini07} found that $L_X > 10^{42}$ \ergs\ [0.5--8 keV] 
AGN were more centrally concentrated than typical cluster galaxies, while 
AGN an order of magnitude less luminous had the same distribution as the 
inactive galaxy population. For both luminosity thresholds the velocity 
distribution of the AGN were consistent with the galaxy population. 

It is more challenging to compare these higher-luminosity X-ray AGN to the host 
galaxy population because we lack membership data for nearly all of the 
high-redshift clusters. Nevertheless, we can compare the distribution of 
sources to the typical distribution of cluster galaxies and to the excess 
surface density distribution found by surveys of X-ray point sources toward 
distant clusters.  In Figure~\ref{fig:rad} we present a histogram of the number 
of X-ray AGN from the cluster center as a function of distance in both physical 
units (Mpc) and normalized to $R_{200}$. While the sample is small, two results 
are apparent from the figure. First, there are approximately equal numbers of 
AGN outside $0.5 R_{200}$ as inside it, whereas if the AGN traced the cluster 
galaxy distribution we would expect them to be more centrally concentrated. 
Second, the radial distribution is more strongly peaked when plotted in 
physical units than normalized to $R_{200}$; 

While we do not have detailed information on the radial distribution of the 
cluster galaxy populations in these clusters, we do have extensive data 
on nearby clusters from \citet{christlein03}. For these clusters we have 
investigated the cluster galaxy distribution with the same selection 
criteria ($R<R_{200}$, $M_R<M_R^*+1$, $\Delta v<3\sigma$) and find that 
70\% of the galaxies fall within $0.5 R_{200}$, whereas 10 of 18 luminous 
AGN at $z>0.4$ are 
within $0.5 R_{200}$. The binomial probability is only 14\% that we would 
find 10 or fewer AGN within $0.5 R_{200}$ if we expected 70\%. There is 
thus a mild tendency for luminous AGN to be distributed toward the outskirts of 
the clusters, although this does make the substantial assumption that 
the radial distribution of galaxies within clusters is similar at $z\sim0.8$ 
and the present. This broad distribution in radius is in contrast to our 
earlier results on lower-luminosity AGN in lower-redshift clusters. At low 
redshift we found that 50\% of the luminous AGN were within $0.1 R_{200}$
\citep{martini07}. Better statistics could determine if the AGN are 
preferentially located in the outskirts of clusters compared to all cluster 
galaxies. That would be consistent with the hypothesis that AGN are triggered 
by mergers during infall. From simulations \citet{ghigna98} find that mergers 
between galaxies do not occur within the virial radius. We note that 
\citet{berrier09} simulated the formation of 53 galaxy clusters and find 
most cluster galaxies do not experience `preprocessing' in group environments 
and therefore processes specific to clusters must largely be responsible for 
the differences between cluster and field galaxies. 

The second result has interesting implications for studies that use the 
surface density distribution of excess sources to characterize the 
distribution of AGN in clusters \citep{ruderman05,gilmour09,galametz09}. 
These studies generally plot the excess surface density as a function of 
physical distance from the cluster center and find a central peak in surface 
density. Our results indicate that the true distribution may be flatter 
than implied by use of the physical (proper) distance from the cluster 
core. This is because those surveys, like the present study, include 
clusters with a wide range of masses and consequently a wide range of 
$R_{200}$. Simply adding the distributions for all clusters without 
renormalizing each observation for the size of the cluster will produce an 
artificial central peak due to the mass range of the cluster sample.


\begin{figure}
\plotone{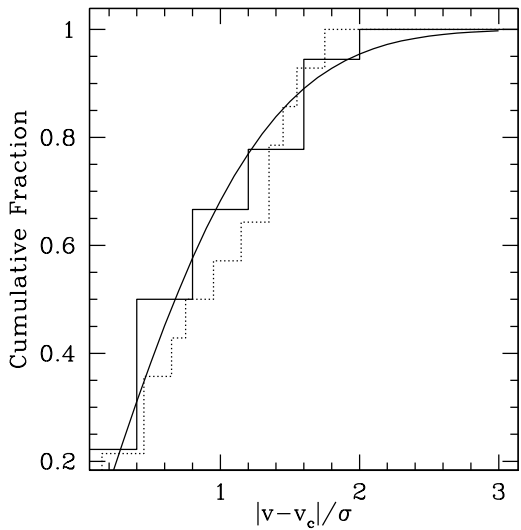} 
\caption{
Histogram of the cumulative velocity distribution of cluster AGN normalized 
to the cluster velocity dispersion for the 18 cluster AGN with $z>0.4$ 
({\it solid histogram}). The AGN velocity distribution is consistent with a 
Gaussian distribution ({\it solid curve}) and the $L_{X,B} \geq 10^{42}$ 
\ergs\ AGN from \citet{martini07} ({\it dotted histogram}). 
\label{fig:vel}
}
\end{figure}

If the cluster AGN are associated with a population that recently entered 
the cluster potential, the host galaxies may also be preferentially on more 
radial orbits and have a larger velocity dispersion than that 
of all cluster galaxies. As noted previously, this is true of 
the emission-line galaxy population in clusters. In Figure~\ref{fig:vel} we 
plot the cumulative velocity distribution for all 18 AGN with $z>0.4$ 
normalized by the cluster velocity dispersion. The distribution is 
in excellent agreement with a Gaussian distribution and we therefore 
find no evidence that the cluster AGN have a larger velocity distribution that 
would be consistent with more radial orbits. This was also found for the 
14 relatively luminous ($L_{X,B} \geq 10^{42}$ \ergs) AGN studied by 
\citet{martini07}. A better test would be to 
compare the AGN host population to the absorption-line galaxies in the 
clusters since the velocity dispersion estimates for many of these clusters 
may be biased toward the emission-line galaxy population because it is 
easier to measure redshifts for them. While this is not the case for 
those whose velocity dispersions are estimated from X-ray data, it may also be 
true of the calibration sample for the relations between X-ray properties 
and galaxy velocity dispersion. 

\subsection{Luminosity Function} \label{sec:xlf} 


\begin{figure}
\plotone{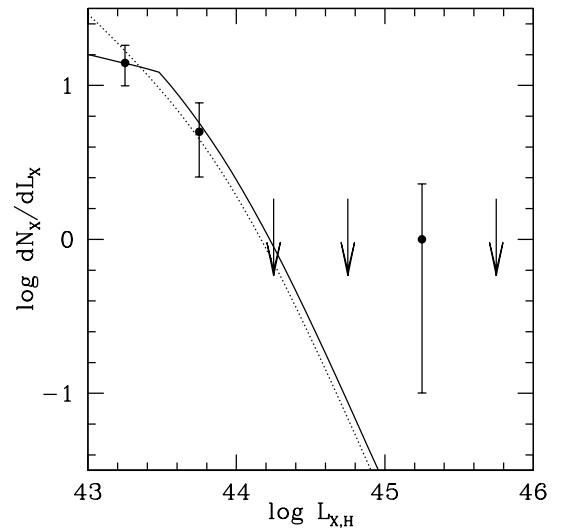} 
\caption{
\label{fig:hxlf} 
Hard X-ray luminosity function of cluster AGN at $z>0.4$ compared to the field 
XLF from \citet{ueda03} at the median cluster redshift ($z=0.8$, {\it solid 
curve}) and at low redshift ($z=0.1$, {\it dotted curve}). The field XLFs have 
been renormalized to be consistent with the cluster measurements in the first 
two luminosity bins. The arrows are upper limits calculated with Poisson 
statistics. 
}
\end{figure}

We have begun to acquire sufficient numbers of cluster AGN that it is 
possible to compare the X-ray luminosity function (XLF) between clusters and 
the field, as well as the cluster XLF at different redshifts. A comparison 
between the cluster and field XLF is interesting because differences 
between the two would be a signature of environment-dependent downsizing. 
There is evidence that this is true of star formation in different 
environments. For example \citet{kauffmann04} find that substantial star 
formation is only present in higher mass galaxies in lower density 
environments in the local universe. 
If the cluster black holes primarily grew at higher redshifts 
than field black holes, similar to the earlier formation epoch expected for 
the stellar populations in luminous cluster galaxies, then the cluster 
luminosity function at high-redshift may have a similar shape to the 
present-day XLF in the field. One test of this hypothesis is to compare the 
characteristic luminosity $L_X^*$ between clusters and the field. If the 
cluster AGN primarily grew at an earlier epoch, $L_X^*$ would be smaller in 
clusters relative to the field at a given redshift. 

It is reasonable to compare the shape of the XLF between clusters and the 
field because the XLF is a measurement of the X-ray sources alone within 
well-defined volumes, although the caveats associated with large scale 
structure discussed in \S\ref{sec:lss} do apply. This is different from the 
case in Section~\ref{sec:evolve}, where we noted that the comparison of 
the evolution of the AGN fraction and the integrated space density was not 
comparing identical 
quantities because the AGN fraction includes information about the galaxy 
population.  The one assumption that we do make is that all of the X-ray 
sources are hosted by galaxies above our threshold, but this is reasonable 
given Figure~\ref{fig:mr}. In addition, the normalization remains arbitrary 
because it is challenging to define a total volume for the cluster AGN sample, 
although this is not necessary because the shape of the XLF already provides 
useful information. In Figure~\ref{fig:hxlf} we plot the cluster XLF for 
our $z>0.4$ sample compared to the field XLF at the median cluster redshift 
of $z=0.8$ from \citet{ueda03}. The cluster XLF is in reasonable agreement 
with the field XLF at the same redshift, although the statistics are quite 
limited. As motivation for future work, we also plot the field XLF at lower 
redshift ($z=0.1$, {\it dotted line}). For the lower-redshift XLF $L_X^*$ is 
smaller and consequently all $L_{X,H} \geq 10^{43}$ AGN are above the 
characteristic luminosity, while these data straddle $L_X^*$ in the field 
XLF at $z=0.8$. Improved statistics for cluster X-ray AGN at $z>0.4$ could 
determine if there is also a break in the cluster XLF, or if it is more similar 
to the field XLF at lower redshift. 

The evolution of the cluster XLF with redshift is also relevant for the 
origin of X-ray AGN in lower-redshift clusters. If cluster AGN at the present 
day are simply the descendants of AGN at higher redshift that have been fading 
for several Gyr, then the difference between the low-redshift and 
high-redshift cluster XLF should be consistent with pure luminosity evolution. 
In contrast, if there is substantial retriggering of low-luminosity AGN in 
low-redshift clusters, or if other mechanisms are capable of fueling AGN in 
clusters, then the cluster XLF evolution may not be consistent with just 
luminosity evolution. A signature of other fueling or triggering mechanisms 
would be a substantially larger population of lower-luminosity AGN in present 
day clusters compared to expectations from the high-redshift population. 
While pure luminosity evolution would be surprising because this is not 
observed in field AGN, the most luminous cluster galaxies are consistent with 
passive evolution. Better measurements of the cluster XLF over a broader range 
in luminosity could investigate this hypothesis. 

\subsection{Host Galaxy Properties} \label{sec:hosts} 

Both the colors and morphologies of low-luminosity ($\sim10^{41}$ \ergs) AGN 
in low-redshift clusters suggest they are primarily hosted by galaxies 
dominated by light from their old stellar populations \citep{martini06}. 
This becomes progressively less true for higher-luminosity AGN and 
ground-based observations of the most luminous sources ($\geq 10^{43}$ erg/s) 
in Abell 2104 \citep{martini02} and Abell 754 \citep{arnold09} indicate 
that they have late-type morphologies, although their hosts are luminous. 
In addition, these more luminous AGN are more likely to exhibit 
visible-wavelength AGN 
spectral signatures than their lower-luminosity counterparts. 

While the spectroscopic classification of the high-redshift sample is fairly 
subjective because of variations in wavelength coverage and signal to noise
ratio, the spectroscopic classifications reported by \citet{silverman05a} 
and \citet{eckart06} support the low-redshift results. They classified six 
of 17 X-ray AGN as BLAGN, nine as other emission-line galaxies, and the 
remaining two as absorption-line galaxies. The vast majority of the higher
luminosity AGN have substantial line emission, even with the bias against 
redshift measurements for sources without strong emission lines. We note 
for comparison that two of the six AGN in the large-scale structure sample 
are classified as BLAGN and the other four are evenly split between 
emission-line and absorption-line galaxies. These other sources are thus 
similar to the cluster AGN. 

Several of the high-redshift clusters also have HST observations suitable 
to study the morphologies of the cluster galaxies. The largest survey of 
X-ray source morphology in high-redshift clusters is that by \citet{martel07}, 
who investigate the fields of five high-redshift clusters: RX J0152-1357, 
RX J0849+4452, RDCS J0910+5422, MS 1054-0321, and RDCS J1252-2927, and the 
middle three clusters overlap this sample. For the entire field sample they 
classify half of the X-ray counterparts as early-type, 35\% as late-type, 
and 15\% as irregular galaxies. For the six cluster members in their sample, 
they find half are in early-type hosts, two in late-type hosts, and one in 
an irregular galaxy. In addition, three of these cluster AGN hosts are 
in interacting systems. The specific overlap with this sample are 
CXOU J091043.3+542152 and CXOMP J105650.6-033508 and both have early-type 
morphologies (their other member in RDCS J0910+5422 falls slightly below 
our luminosity threshold). 

\subsection{Implications for the Sunyaev-Zel'dovich Effect} \label{sec:sz} 

Many cluster surveys are currently planned or in progress that use the 
Sunyaev-Zel'dovich effect to identify large numbers of clusters 
\citep[e.g.][]{kosowsky03,ruhl04}. This effect is caused by inverse Compton 
scattering of Cosmic Microwave Background (CMB) photons off hot electrons in 
the ICM that changes the spectrum of the CMB in the direction of a cluster
\citep[e.g.][]{carlstrom02}. The main virtue of this effect is that it is 
redshift independent, and consequently can be used to detect (the hot 
electrons associated with) clusters out to high redshifts. However, mechanical 
heating by AGN in the cluster may contribute to the thermal energy of 
the ICM \citep[e.g.][]{birzan04} and thus make it more difficult to 
identify some clusters. Any increase in the AGN population with redshift will 
also introduce a systematic effect with redshift. 

The potential impact of AGN on SZE cluster surveys was recently examined in 
detail by \citet{lin07}. They measured the radio luminosity function in 
nearby clusters at 1.4 GHz and used measurements of AGN at higher frequencies 
\citep{cooray98,coble07} to estimate that on order 10\% of clusters will 
have an AGN flux comparable to the SZE flux. As a worst-case scenario 
they adopted an evolution model where the fraction of radio AGN increases 
as $(1+z)^{2.5}$. This model was largely motivated by observations of the 
radio galaxy luminosity function, which suggested evidence for an increase
\citep{best02,branchesi06}. If this population evolves at a comparable rate
more consistent with the $(1+z)^{5.3}$ rate we observe for luminous X-ray AGN,
then the fraction of substantially contaminated clusters will be 
higher than predicted by \citet{lin07}. 

\section{Discussion} \label{sec:dis} 


The extent of the correlation between the evolution of star formation and AGN 
in clusters could provide valuable new insights into how closely related these 
two processes are. The original work by \citet{butcher78,butcher84} on the 
evolution of the fraction of blue galaxies in clusters provides a useful first 
point of comparison to the AGN fraction evolution, in part because we 
adopted many elements of their methodology. Specifically, \citet{butcher84} 
characterized 
cluster galaxy evolution with: 1) a fixed criterion to define the sample of 
interest (a galaxy was classified as blue if the rest-frame $B-V$ color was 
at least 0.2 mag bluer than the relation exhibited by the red galaxies); 2) 
measurement of this population relative only to cluster galaxies above some 
luminosity threshold ($M_V = - 20$); 3) use of an aperture scaled to the 
physical properties of individual clusters (a circle that contained the inner 
30\% of the cluster galaxy population). With these definitions, 
\citet{butcher84} found that the blue galaxy fraction increased from 
$f_B ~\sim 0.03$ at $z\leq0.1$ to $f_B \sim 0.25$ at $z=0.5$ for relatively 
compact, concentrated clusters, or approximately an order of magnitude. 

One of the most recent and comprehensive studies of the evolution of 
star formation in clusters is the work of \citet{poggianti06}. These
authors used the \oii $\lambda 3727$ line as a tracer of star formation, rather 
than color, and measured the fraction of galaxies with \oii\ emission 
(equivalent width $> 3$\AA) as a function of both cluster redshift 
and cluster velocity dispersion. Their sample includes 25 clusters with 
$z = 0.4 - 0.8$ and another 10 groups in the same redshift range, while they 
have a large local comparison sample at $z=0.04 - 0.08$ from the Sloan Digital 
Sky Survey. They measure the \oii\ fraction $f_{{\rm [OII]}}$ relative to an 
evolving absolute magnitude limit $M_{V,lim}$ that varies from -20.5 at $z = 
0.8$ to -20.1 at $z = 0.4$, while the local limit was $M_V < -19.8$. Their 
main results are that there is substantial evolution in $f_{{\rm [OII]}}$ and 
that there is substantial variation in $f_{{\rm [OII]}}$ with velocity 
dispersion at a given redshift. 
Given the velocity dispersion dependence, a 
direct comparison of the evolution of $f_{{\rm [OII]}}$ with $f_A$ is not 
meaningful for different cluster samples. Instead, we have used their upper 
envelope for 
$f_{{\rm [OII]}}$($\sigma$) at high redshift and their envelope prescription 
at low redshift to estimate $f_{{\rm [OII]}}$ for each of our clusters and 
then computed the average $f_{{\rm [OII]}}$ for each of the subsamples shown 
in Table~\ref{tbl:fabin}. These relations predict an increase in 
$f_{{\rm [OII]}}$ of less than a factor of two from the low-redshift 
to the high-redshift subsamples, or substantially less than the factor of 
eight we observe for the AGN fraction. 

These results are interesting, although numerous caveats forestall 
too much interpretation of the relative rates of evolution. One major 
concern is that there is likely downsizing in clusters similar to what is 
observed in the field \citep[e.g.][]{cowie96,hasinger05,silverman08,yencho09}, 
that is the relative number of galaxies with star formation or AGN activity 
above a certain threshold varies with redshift. The direct implication of 
this for the AGN fraction is that the evolution of the AGN fraction 
over a given redshift range is expected to depend on luminosity, just 
as the rate of evolution of the AGN space density is observed to vary in 
the field as a function of minimum luminosity \citep[e.g.][]{ueda03}. 
This is similarly a complication for interpretation of the evolution of 
star formation, and consequently limits direct comparison of the mere rates of 
evolution of star formation and AGN above some threshold. For example, while 
\citet{poggianti06} have similarly used an evolving galaxy luminosity threshold 
to characterize the evolution of the star-forming galaxy fraction, their 
galaxy luminosity threshold is over a magnitude fainter and therefore they 
have measured the evolution in a population that includes many more fainter 
cluster members.
However, these concerns are not an obvious limitation to comparisons that use 
the same luminosity threshold to separately compare either AGN or star 
formation across different environments, particularly when the evolution of the 
star formation rate and AGN luminosity are tied to the same galaxy population. 
For example, if the relative rates of evolution of AGN and star formation in 
$< M_R^*+1$ galaxies were different in the field and clusters, this would 
suggest a limit to the extent of the apparent coevolution of black holes and 
galaxies in at least one of these environments. 

Another concern about a direct comparison to these measurements of the 
evolution of the star forming galaxy population is that \oii\ emission 
is more susceptible to reddening and metallicity effects relative to other 
star formation indicators, such as H$\alpha$ \citep{kewley04}. Many 
ISO studies \citep[summarized by][]{metcalfe05} found evidence for an 
increase in star formation in clusters at higher redshifts, and that 
the increase appeared to be greater than that predicted by UV continuum or 
visible-wavelength spectroscopic diagnostics. \spitzer\ observations of
clusters have also found substantial, often obscured, star formation 
in high-redshift clusters \citep{geach06,marcillac07,bai07}. \citet{geach06} 
used new \spitzer\ data for two clusters and data for five others  
from the literature to estimate the star formation rate 
normalized by the cluster mass. They find evidence for an increase in 
higher redshift clusters, but also substantial variation between clusters 
at the same redshift. \citet{saintonge08} used a larger sample of eight 
clusters with 24$\mu$m \spitzer\ data to study the evolution of the fraction of 
obscured star-forming galaxies from $z=0.02 - 0.83$. 
They find that the fraction of cluster galaxies with star formation rates 
above 5 \msun yr$^{-1}$ increases from 3\% at $z=0.02$ to 13\% at $z=0.83$ 
and that this is stronger evolution than exhibited by color-selection, such 
as the criteria of \citet{butcher78,butcher84}. The star-forming galaxies they 
identify in 
these clusters are also mostly disjoint from the Butcher-Oemler galaxies and 
consequently when they sum the blue and mid-infrared galaxies the fraction of 
star-forming galaxies increases to $\sim 23$\% at high redshift. 

Several of these \spitzer\ studies overlap clusters that are also in our 
sample and it is interesting to see if there is a direct correspondence 
between the AGN and mid-infrared sources detected by \spitzer. 
The massive cluster MS1054-03 was studied by \citet{bai07} and their 24$\mu$m 
sources include the two X-ray AGN identified by \citet{johnson03}. 
\citet{saintonge08} have three clusters in common with our sample: 
MS0451.6-0305, MS2053.7-0449, and MS 1054-03, although they do not provide 
information on individual sources. While not in our sample, the study of 
RX J0152.7-1357 ($z=0.831$) by \citet{marcillac07} found that the two most 
luminous $24\um$ sources (of 22 confirmed members) were also X-ray AGN. 
Similarly, \citet{geach09} found that one (of 12) of the luminous 
infrared galaxies ($L_{IR} > 10^{11} L_\odot$) in CL0024+16 ($z=0.4$) 
was obviously an AGN based on their infrared data alone. At lower 
redshifts, \citet{gallagher08} have also used \spitzer\ data to 
identify AGN and star forming galaxies in Hickson Compact Groups. 

\citet{saintonge08} explore whether or not the increase in the fraction of 
obscured star formation in high-redshift clusters is related to infall. 
They speculate that the increase in star formation reflects the infall of 
new members and note that most of the MIPS-detected cluster galaxies are 
not projected onto the cluster core (inner 500pc). Over larger scales the 
work of \citet{gallazzi09} explored the obscured star formation fraction 
as a function of environment in the Abell 901/902 supercluster at $z=0.165$. 
They find more obscured star formation at intermediate densities than in the 
cluster cores, similar to the distribution of the AGN population studied 
by \citet{gilmour07} in the same supercluster. 
If there is a substantial increase in the obscured star 
formation fraction in the intermediate densities around clusters, and 
the star formation in this environment increases with redshift, then 
projection of some of these structures onto the cluster core may 
contaminate the cluster estimates. 


As discussed in Section~\ref{sec:lss}, AGN in the large-scale environments 
around massive clusters may also project onto cluster cores. To better 
evaluate this possibility, it is useful to both directly measure the 
AGN population immediately outside clusters and measure the AGN population 
in intermediate densities more generally. Just as \citet{poggianti06} found 
that the fraction of \oii-emitting galaxies increases in lower velocity 
dispersion environments, the AGN fraction as a function of environment is 
important because the environmental dependence may provide new information 
on the processes that drive AGN evolution. Both the XMM observations of the 
COSMOS fields \citep{silverman09a,silverman09b} and \chandra\ observations of 
the Extended Groth Strip from DEEP2 \citep[the All-wavelength Extended Groth 
strip International Survey, AEGIS;][]{georgakakis08a,georgakakis08b} have 
estimated the AGN fraction in groups of galaxies or as a function of 
local overdensity at high redshifts. \citet{georgakakis08b} found that X-ray 
AGN are more frequently found in groups than in the field, which they 
connect to their observation that the X-ray AGN host galaxies are often 
red, luminous galaxies that tend to reside in denser environments, although 
they also find that this trend may reverse for the most powerful AGN.
In a narrower redshift range from $0.7 < z < 0.9$ and for $M_B < -20$ mag 
they find that the AGN fraction is comparable in groups and the field and 
about 5\%. This is approximately a factor of five higher than we find in 
clusters at similar redshifts, although these values are not exactly 
comparable as the \citet{georgakakis08b} AGN include somewhat lower-luminosity 
sources than our sample and the host galaxy magnitude limits are somewhat 
different. \citet{silverman09a} also investigate the environment dependence of 
X-ray AGN hosted by galaxies above a fixed stellar mass and find no strong 
preference between the field and groups except for the most massive galaxies, 
while \citet{jeltema07} find that the fraction of \oii-emitting galaxies in 
intermediate-redshift, X-ray-selected groups ($0.2 < z < 0.6$) is similar to 
clusters at the same redshift. 

The clustering analysis by \citet{coil09} on the AEGIS data also helps to 
elucidate the distribution of AGN at high redshift as a function of 
environment, AGN luminosity, and host galaxy mass. They find that the X-ray 
AGN have similar clustering to luminous red galaxies and are more likely to 
reside in groups, while UV-bright QSOs are less strongly clustered and more 
similar to the field blue galaxy population. This is also similar to the 
results from \citet{kauffmann04} at low redshifts from SDSS, who find that 
galaxies at a fixed stellar mass that host luminous \oiii\ emission are twice 
as common in low-density regions as high. Taken together, the AEGIS and COSMOS 
results illustrate that the measured AGN fraction depends on both the stellar 
mass (or luminosity) of the galaxy population and the star formation rate of 
the host, in addition to the AGN luminosity. This makes a direct comparison 
between these two surveys, as well as to our work on high-redshift clusters, 
somewhat problematic. The X-ray range considered by \citet{silverman09a} 
extends over $42 < {\rm log} L_{0.5-10 {\rm \, keV}} < 43.7$, or approximately 
half an order of magnitude below our X-ray threshold for a typical AGN SED. 
The X-ray AGN studied by \citet{coil09} extend an order of magnitude fainter 
than our work to a hard band limit of $L_{X,H} > 10^{42}$ \ergs. Both of these 
surveys are therefore dominated by intrinsically less luminous objects. The 
galaxy mass and luminosity ranges are similarly not identical. In future work 
we hope to put all of these high-redshift measurements on an equal basis for a 
more direct comparison. 

While none of these results suggest that there are more luminous AGN in 
clusters than groups or the fields out to $z\sim1$, such a trend may be 
seen at yet higher redshits. Observations of cluster galaxies, particularly 
massive cluster ellipticals, suggest that most of their stars formed 
earlier than field galaxies \citep[by 0.4 Gyr;][]{vandokkum07}. If the central 
black holes of these galaxies grew contemporaneously, then perhaps by 
$z\sim2$ the AGN fraction will be higher in denser environments. Some 
interesting support for this picture comes from \chandra\ observations of the 
SSA22 protocluster at $z=3.09$ \citep{lehmer09}. They find a slightly higher 
AGN fraction in Lyman Break and Ly$\alpha-$emitting galaxies in the 
protocluster compared to the field. While this is just one region, 
observations of the AGN fraction in clusters relative to the field at $z\sim2$ 
and above could provide interesting new insights into the coevolution of black 
holes and galaxies. 

\section{Summary} \label{sec:conclude} 

We have conducted an expanded survey to identify luminous $L_{X,H} \geq 
10^{43}$ \ergs\ AGN in clusters of galaxies from $z \sim 0.05$ to $z \sim 1.3$. 
At low redshifts we have presented a new X-ray analysis of archival \chandra\ 
observations and spectroscopic follow-up of AGN candidates in six new clusters. 
There are no new, luminous AGN in these clusters and there are a total of just 
two luminous AGN in our sample of 17 clusters with $z<0.4$. These measurements
further strengthen the evidence for a very small luminous AGN fraction in 
low-redshift clusters. An important virtue of the new clusters is that the 
X-ray and spectroscopic coverage extends to the projected $R_{200}$ radius and 
therefore they are better matched to observations of high-redshift clusters. 
At higher redshifts we have combined our previous work with literature data on 
X-ray sources, primarily from the ChaMP and SEXSI surveys, to compile a 
total sample of 15 clusters at $z>0.4$. In spite of somewhat incomplete 
spectroscopic coverage of the X-ray sources in these fields, there are 18 
luminous AGN in these clusters.  

We parameterize the evolution of the AGN population in clusters in terms of 
the fraction of luminous galaxies that host AGN above our luminosity threshold. 
We have used a variety of techniques to estimate the number of luminous 
galaxies, defined to have $M_R < M_R^*+1$, in these clusters and calculated 
the average cluster AGN fraction in several redshift bins. As the low and 
high-redshift clusters are reasonably well matched in terms of cluster 
velocity dispersion and X-ray temperature, the increase in the number of 
AGN is closely related to the increase in the fraction of galaxies 
more luminous than $M_R^*+1$. Specifically, we find that the AGN fraction 
increases by approximately a factor of eight from $z\sim 0.2$ to $z \sim 1$. 
This corresponds to an increase in the AGN population that scales as 
$(1+z)^{5.3}$. If the radio AGN population in clusters increases by a 
comparable amount, radio AGN may impact the identification of clusters as a 
function of redshift in current and planned SZ surveys. The substantial 
evolution in the cluster AGN population is also correlated with the evolution 
of the fraction of star-forming galaxies in clusters known as the 
Butcher-Oemler effect. Detailed studies of star formation and AGN in 
individual clusters could better quantify the extent that these two phenomena 
are coupled in clusters or perhaps even individual galaxies. We have also 
estimated the evolution of the field AGN fraction to compare it to the cluster 
AGN fraction. While the field AGN fraction is higher at all redshifts, the 
present data do not suffice to conclude if the rate of evolution is faster 
or slower in clusters. Future measurements of the relative evolution of star 
formation and black hole growth in clusters and the field could be an 
important probe of the coevolution of black holes and their host galaxies. 

Measurements of the radial distribution of the cluster AGN provide new 
information on the origin of AGN within clusters. Unlike we found in previous 
work at low redshifts, the AGN in these high-redshift clusters are not strongly 
centrally concentrated when their distribution is plotted normalized to the 
$R_{200}$ radius. This demonstrates that there are substantial numbers in the 
outskirts of clusters and supports the hypothesis that some cluster AGN are 
hosted by relatively gas-rich galaxies that have recently entered the cluster 
potential. While this excess is not apparent in the velocity distribution, 
this may be due to biases in the measurement of the cluster velocity dispersion 
or simply small number statistics. We have also presented the first measurement 
of the XLF of cluster AGN at high-redshift and found that it is consistent with 
the field XLF at the same redshift. This comparison illustrates the future 
potential of XLF measurements in clusters to measure environment-dependent 
downsizing in clusters, as well as how the evolution of the cluster XLF can 
be used to constrain the evolution of black hole growth in clusters. 

\acknowledgements 

We are grateful to John Silverman and the referee for many suggestions that 
have improved this paper. We also acknowledge helpful discussions with Dan 
Stern and Tommaso Treu. Support for this work was provided by the National 
Aeronautics and Space Administration through Chandra Award Number AR8-9014X 
issued by the Chandra X-ray Observatory Center, which is operated by the 
Smithsonian Astrophysical Observatory for and on behalf of the National 
Aeronautics Space Administration under contract NAS8-03060. 
PM is grateful for support from the NSF via award AST-0705170 and from the 
Department of Astronomy at The Ohio State University. 
This research has made use of the NASA/IPAC Extragalactic Database (NED) 
which is operated by the Jet Propulsion Laboratory, California Institute of 
Technology, under contract with the National Aeronautics and Space 
Administration.

{\it Facilities:} \facility{Hiltner (), CXO ()}


\end{document}